\documentclass[aps,prb,amsmath,amssymb,twocolumn, superscriptaddress]{revtex4-2}

\usepackage{graphicx}
\usepackage{adjustbox}
\usepackage{bm}
\usepackage{color}
\usepackage{braket}
\usepackage{standalone}
\usepackage{multirow}
\usepackage{tikz}
\usepackage{mathrsfs}
\usepackage[utf8]{inputenc}
\usepackage{dsfont}
\usepackage[colorlinks,bookmarks=true,citecolor=blue,linkcolor=red,urlcolor=blue]{hyperref}

\newcommand{\eq}[1]{Eq.~(\ref{#1})}
\newcommand{\affTUD}{Institute of Theoretical Physics${\rm ,}$ Technische Universit\"{a}t Dresden and W\"{u}rzburg-Dresden Cluster of Excellence ct.qmat${\rm ,}$ 01062 Dresden${\rm ,}$ Germany}
\newcommand{\affMPIKS}{Max Planck Institute for the Physics of Complex Systems, N\"{o}thnitzer Str. 38, 01187 Dresden, Germany}

\usepackage{bbm}
\usepackage[subrefformat = parens, caption = false, labelformat=parens]{subfig}

\usepackage{floatrow}
\usepackage{verbatim}

\usepackage[acronym]{glossaries}
\makenoidxglossaries

\newacronym{iqh}{IQH}{integer quantum Hall}
\newacronym{pbc}{PBC}{periodic boundary conditions}
\newacronym{obc}{OBC}{open boundary conditions}
\newacronym{dos}{DOS}{density of states}
\newacronym{gf}{GF}{Green's function}
\newacronym{kpm}{KPM}{kernel polynomial method}
\newacronym{wrt}{w.r.t.}{with respect to}

\glsdisablehyper

\begin{document}

\title{Quantum Hall Effect without Chern Bands}
\author{Benjamin Michen}
\email{benjamin.michen@tu-dresden.de}
\affiliation{\affTUD}
\author{Jan Carl Budich}
\email{jan.budich@tu-dresden.de}
\affiliation{\affTUD}
\affiliation{\affMPIKS}

\date{\today}

\begin{abstract}
The quantum Hall effect was originally observed in a two-dimensional electron gas forming Landau levels when exposed to a strong perpendicular magnetic field, and has later been generalized to Chern insulators without net magnetization. Here, further extending the realm of the quantum Hall effect, we report on the robust occurrence of an integer quantized transverse conductance at the onset of disorder in a microscopic lattice model all bands of which are topologically trivial (zero Chern number). We attribute this phenomenon to the energetic separation of non-quantized Berry fluxes within those bands. Adding disorder then nudges the system into a quantum Hall phase from an extended critical regime obtained by placing the Fermi energy within a broad window inside a trivial band. This natural integer-rounding mechanism manifests as the mobility-gap-induced quantization of a non-universal Hall conductance. Our results are corroborated by numerical transport simulations and the analysis of two complementary topological markers.
\end{abstract}

\maketitle

The quantum Hall effect, one of the most remarkable phenomena occurring in nature, is of foundational importance to the broad research field of topological matter \cite{Klitzing, Prange_IQH, Laughlin_IQH_1, Thouless_1, Halperin_1, Laughlin_IQH_2, Pruisken_1, Pruisken_2, Pruisken_3, Khmelnitskii, Reentrant_QAHE, Scalar_QHE}. Establishing topology as the rationale behind physical robustness, the experimental observation of a transverse conductance $\sigma_{xy}$ quantized in units of $e^2/h$ has been explained with theory in terms of a topological invariant known as the first Chern number \cite{Chern_1946, IQH_Chern_number_1, IQH_Chern_number_2, IQH_Chern_number_3}. This represents a pioneering example of the by now well established paradigm of topological Bloch bands characterized by global properties, the discrete value of which remains unchanged under continuous perturbations \cite{HasanKane2010, Qi2011, Ryu2016, Budich2013}. Adiabatic continuity is in this context largely synonymous with the existence of a finite energy gap that must close at a topological quantum phase transition for a topological property to change \cite{Wen2010, Wen2017}.

\begin{figure}[htp!]	 
{
    \vbox to 0pt {
            \raggedright
            \textcolor{white}{
                \subfloatlabel[1][fig:one_a]
                \subfloatlabel[2][fig:one_b]
            }
        }
}
{\includegraphics[trim={0.75cm 0cm 0.75cm 0.cm}, width=0.95\linewidth]{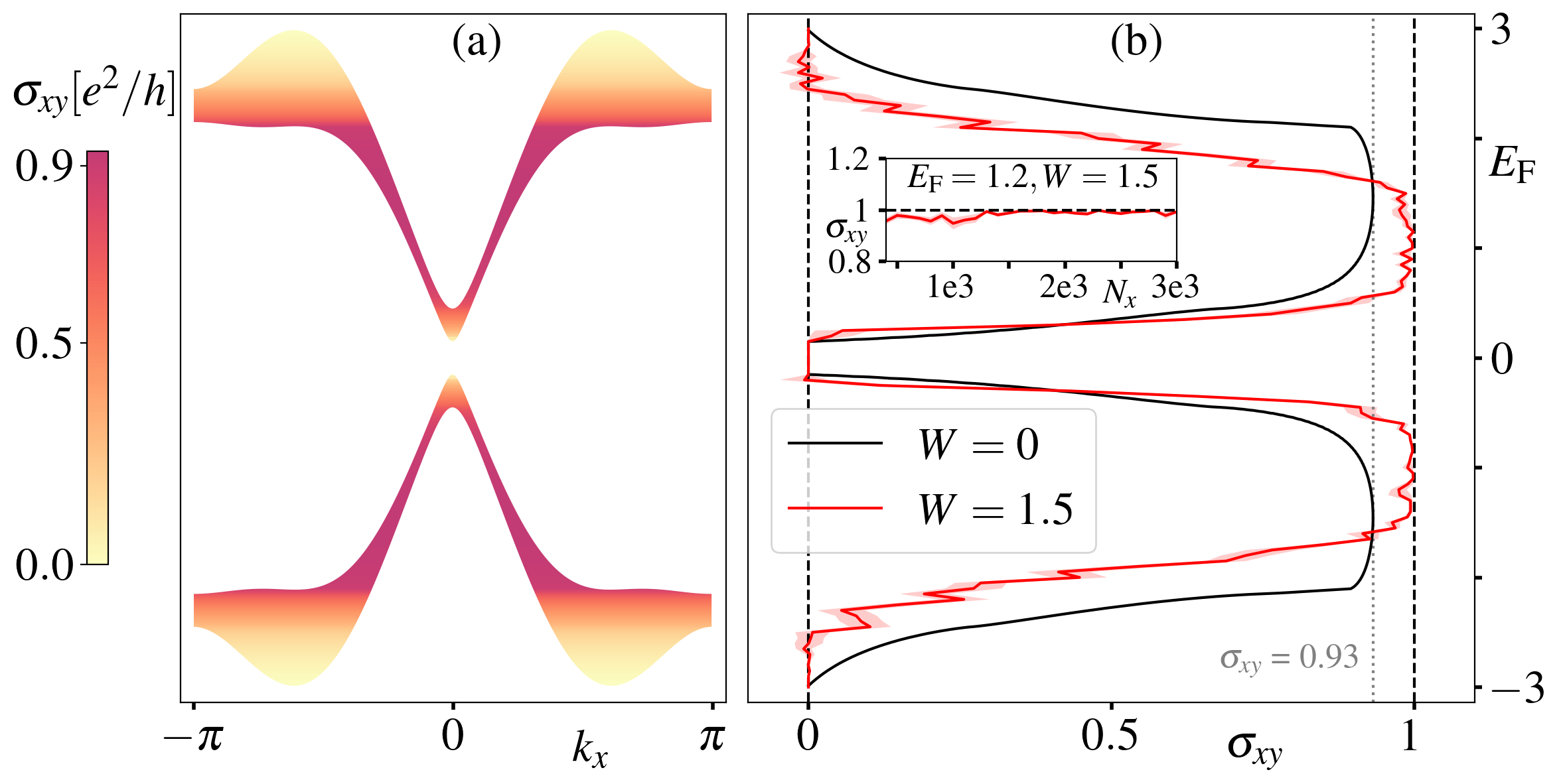}}
\caption{(a) Band structure of the free Hamiltonian $\hat H_0$ [see \eq{eq:ham}]. Color indicates Hall conductance $\sigma_{xy}^{W=0}(E_\mathrm{F})$ of the clean system with Fermi energy $E_\mathrm{F}$, which is proportional to the Berry curvature accumulated up to $E_\mathrm{F}$ [cf.~\eq{Eqn:accumulated_flux}]. (b) $\sigma_{xy}^{W}(E_\mathrm{F})$ at disorder strength $W = 0$ ($W = 1.5$ ) in black (red), and system size $N_x = 2000, N_y = 500$, averaged over $S=40$ disorder realizations, where the shaded corridor indicates the sample standard error. Inset: Finite size scaling at $E_\mathrm{F} = 1.2$ as a function of $N_x$ at fixed $N_x/N_y=4$. Other parameters in all plots are $r = 1.5$, $\epsilon_1 = 0.3$, $\epsilon_2 = 2$, $\gamma  =2$, and $\gamma_2 = 0.3$.
 }\label{fig:one}
\end{figure}

Surprisingly, in this work we provide strong evidence that the topological quantization of the Hall conductance can occur with the onset of disorder in an extended parameter regime of a microscopic lattice model, both Bloch bands of which are topologically trivial. The physical mechanism behind this phenomenology may be seen as a natural integer rounding effect: When placing the Fermi energy $E_\mathrm{F}$ within either band of our model, the non-universal value of $\sigma_{xy}$ is quantized to the closest integer value in units of $e^2/h$ \cite{IQH_Chern_number_2} by the continuous opening of a mobility gap \cite{Anderson_localization, Scaling_theory, Wegner, Wegner, 2D_Anderson_localization, AL_review} upon switching on a on-site disorder potential $\hat W$. The underlying clean model exhibits a wide energy window where $\sigma_{xy} > 0.5 e^2/h$ (see Fig.~\ref{fig:one}) due to the energetic separation of sizable though non-quantized Berry fluxes within its topologically trivial bands. In this light, the clean metallic situation with $E_\mathrm{F}$ inside the eligible window may be seen as an extended critical region that is immediately nudged into a quantum Hall phase by a generic disorder potential.
 
On the other hand, we emphasize that the underlying clean band structure of our model may be continuously deformed into a trivial atomic insulator, because $\sigma_{xy} = 0$ inside the band gap. That way, all structure is removed from the entire microscopic lattice system without a topological quantum phase transition, which clearly makes the phenomena reported in this work quite different from the familiar quantum Hall scenarios based on Chern bands and Landau levels, respectively. As we detail below, our findings also differ qualitatively from several other settings in which topological phenomena have been identified in unconventional relation to Chern numbers \cite{Anomalous_BBC, Anomalous_Floquet_AI, Floquet_levitation, Fine_structure}, in particular from the topological Anderson insulator \cite{TAI, theory_TAI, Mobility_gap_TAI, Disordered_CI_HHH} where a trivial bulk band gap is closed and reopened with increasing disorder strength thus inducing an effective Chern insulator band structure.

\begin{figure}[htp!]	 
{
    \vbox to 0pt {
            \raggedright
            \textcolor{white}{
                \subfloatlabel[1][Fig:Transport_a]
                \subfloatlabel[2][Fig:Transport_b]
            }
        }
}
{\includegraphics[trim={0.75cm 0cm 0.75cm 0.cm}, width=0.95\linewidth]{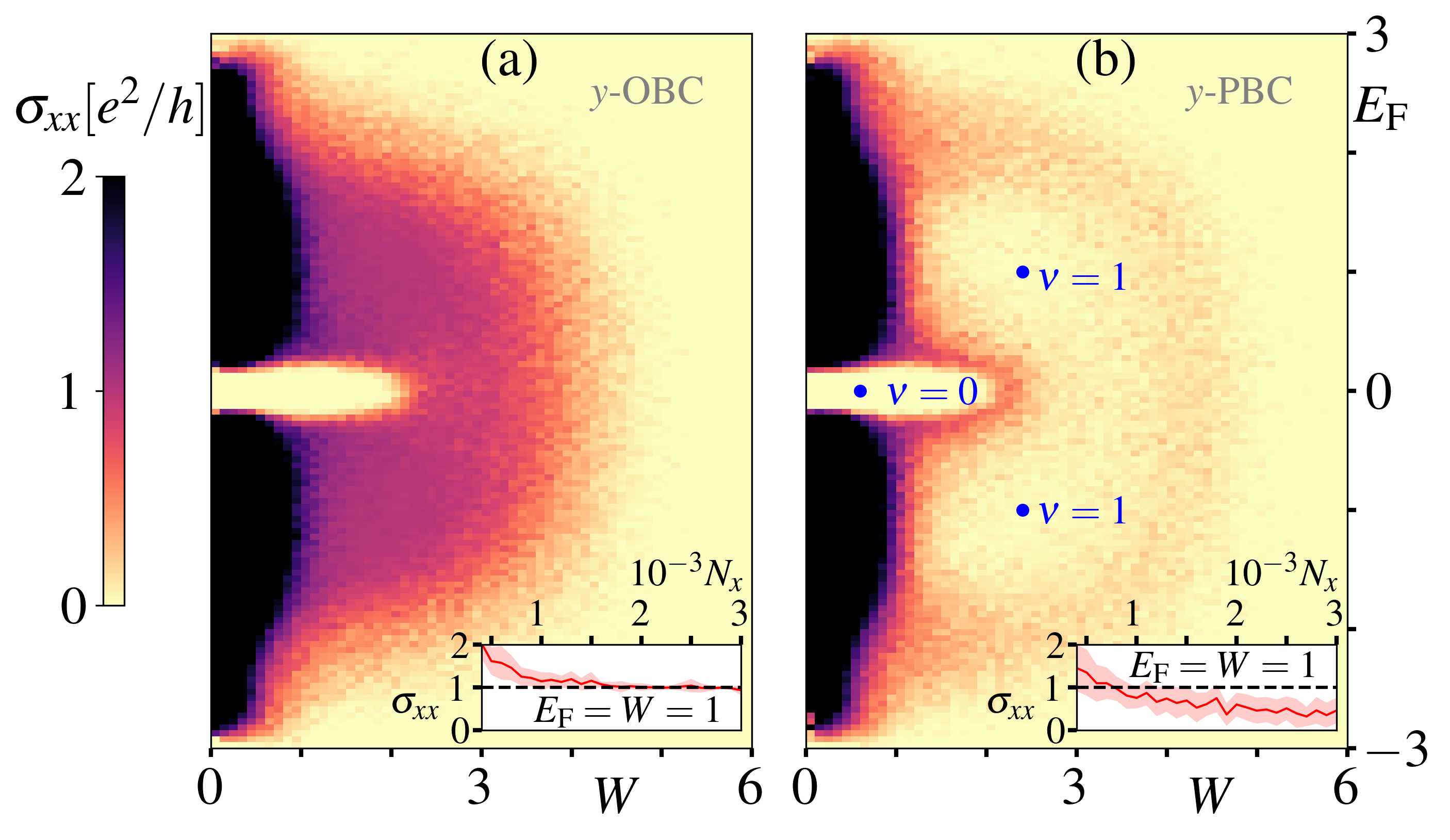}}
\caption{(a) Two-terminal conductance $\sigma_{xx}$ for a system with $N_x = 600$, $N_y = 150$ sites and \gls{obc} in $y$-direction. (b) Same as (a) but with \gls{pbc} in $y$-direction. Winding number $\nu$ [see \eq{Eqn:W_num}] indicated at the blue points in parameter space. Insets: Size scaling of $\sigma_{xx}$ for $E_\mathrm{F} = 1$ and $W = 1$ over $N_x$ at constant $N_x/N_y=4$ with standard deviation indicated as a shaded corridor.  Parameters are $r = 1.5$, $\epsilon_1 = 0.3$, $\epsilon_2 = 2$, $\gamma  =2$, $\gamma_2 = 0.3$, disorder realizations $S=20$.}\label{Fig:Transport}
\end{figure}

\begin{figure}[htp!]	 
{\includegraphics[trim={0.75cm 0cm 0.75cm 0.cm}, width=0.95\linewidth]{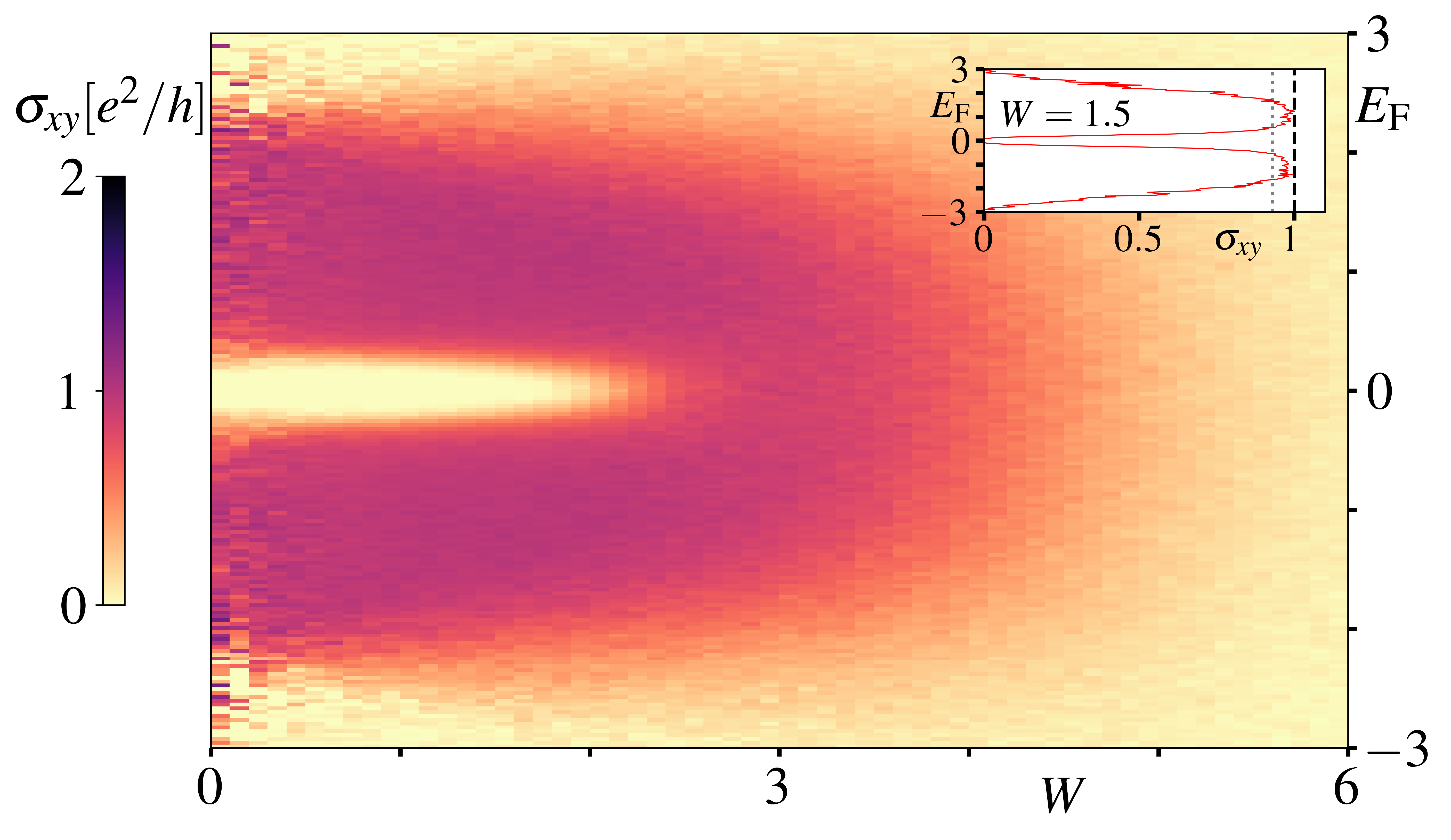}}
\caption{$\sigma_{xy}$ as a function of $W$ and $E_F$ for $N_x = 600$, $N_y = 150$ sites with \gls{pbc}, obtained from the Chebyshev expansion of the Kubo-Bastin formula (\ref{Eqn:Kubo_Bastin}) with $M = 1500$, $R = 5$, and $S = 30$. Parameters are $r = 1.5$, $\epsilon_1 = 0.3$, $\epsilon_2 = 2$, $\gamma  =2$, $\gamma_2 = 0.3$. Inset: Calculation with increased accuracy using $M = 2500$ and $S = 50$ for $W = 1.5$. The maximal clean $\sigma_{xy} \approx 0.93$ is shown as a dotted gray line [cf.  Fig.~\protect\subref*{fig:one_b}].
}\label{Fig:Hall_cond}
\end{figure}

\begin{figure}[htp!]	 
{\includegraphics[trim={0.75cm 0cm 0.75cm 0.cm}, width=0.95\linewidth]{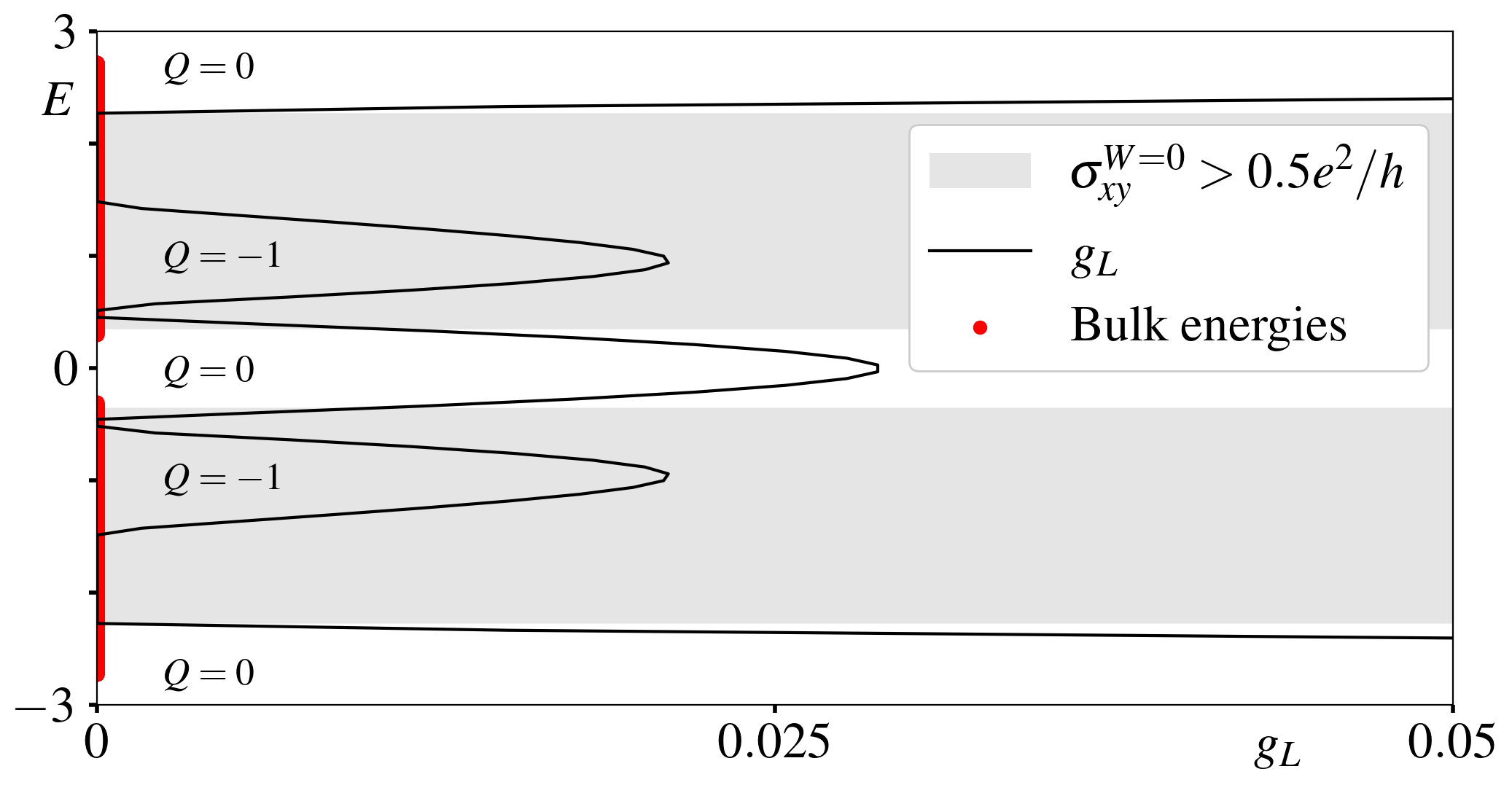}}
\caption{Localizer gap $g_L$ [cf.~\eq{Eqn:loc_gap}] as a function of $E$ for a clean system of size $N_x = 20$, $N_y = 20$ with $\kappa = 0.4$. The shown $g_L$ is the minimum of 100 reference positions inside the central Wigner-Seitz cell. For finite $g_L$, we indicate the value of $Q$ [cf.~\eq{Eqn:loc_index}]. The clean bulk spectrum is shown in red and the energy window where $\sigma_{xy}^{W=0}(E_\mathrm{F})  > 0.5 \,e^2 / h$ [cf.~\eq{Eqn:accumulated_flux}] is marked in gray. Parameters are $r = 1.5$, $\epsilon_1 = 0.3$, $\epsilon_2 = 2$, $\gamma  =2$, $\gamma_2 = 0.3$, $W = 0$.}\label{Fig:spec_loc}
\end{figure}

Besides performing extensive numerical simulations on the transport properties of our model system (see Figs.~\ref{fig:one}-\ref{Fig:Hall_cond}) based on the software package Kwant \cite{Kwant, Kwant_Mumps, KPM_tools}, we verify that a topological invariant defined in the mobility gap \cite{Smatrix_pumping, Smatrix_pumping_2, Scattering_invariants} characterized by a finite localization length of bulk states indeed confirms our predictions regarding the quantization of Hall conductance. A careful analysis of the spectral localizer gap structure \cite{Spec_loc_1, Spec_loc_2, Spec_loc_3, Spec_loc_4, Spec_loc_5, spec_loc_mobility_gap} in the considered system provides further complementary evidence corroborating our results (see Fig.~\ref{Fig:spec_loc}).

{\it Minimal two-banded lattice model. ---}
We introduce a minimal lattice model specified by the Hamiltonian 
\begin{align}
\hat H = \hat H_0 + \hat W,
\label{eq:ham}
\end{align}
where $\hat H_0$ is a free translation-invariant two-band Hamiltonian and $\hat W = W\sum_{j_x, j_y} \sum_{\alpha = a,b} f_{\bm j, \alpha} c_{\bm j, \alpha}^\dagger c_{\bm j, \alpha}$ is the disorder potential with strength $W \ge 0$ and random amplitudes $f_{\bm j, \alpha}$ drawn independently from the uniform distribution on the interval $[-1,1]$. The translation-invariant part is specified by the Bloch Hamiltonian $h_0(\bm k ) = \bm d(\bm k) \cdot \bm \sigma$, with $\bm \sigma$ the vector of Pauli matrices and the coefficient vector $d_x(\bm k) = \gamma \sin(k_x)$, $d_y(\bm k) = \lambda(k_x) \sin(k_y)$, $d_z(\bm k) = \gamma_2[r - \cos(2 k_x)] - \lambda(k_x) \cos(k_y)$, and $\lambda(k_x) =  \epsilon_1 + \epsilon_2 (1 - \cos(k_x))/2 $, which corresponds to a tight-binding model on a two-dimensional (2D) square lattice with unit lattice constant. Containing only nearest-neighbor, plaquette diagonal, and next-nearest-neighbor in $x$-direction hopping terms, our model is local and fully microscopic \cite{Supplemental}. In the following, we will discuss its design principles and exemplify our findings using the parameters $r = 1.5$, $\epsilon_1 = 0.3$, $\epsilon_2 = 2$, $\gamma  =2$, and $\gamma_2 = 0.3$, where both bands are topologically trivial (zero Chern number), and there is a finite band gap [see Fig.~\subref*{fig:one_a}].

{\it Berry flux and transport without disorder. --} To allow a finite transverse conductance, the topologically trivial Bloch bands of our model accommodate non-quantized momentum-local Berry fluxes that, however, compensate each other within each band when integrated over the first Brillouin zone. To get a feeling for the transport properties of the clean system ($W=0$), we express the zero temperature Hall conductance as a function of Fermi energy in terms of the accumulated Berry flux up to the Fermi energy, i.e.
\begin{align}
\sigma_{xy}^{W=0}(E_\mathrm{F}) = \frac{e^2}{h} \frac{1}{2\pi}\int_{E_{\bm k}<E_\mathrm{F}}\mathrm{d}^2k \,  \mathcal F, \label{Eqn:accumulated_flux}
\end{align}
where $\mathcal F$ is the Berry curvature \cite{Supplemental}. The colorcode in Fig.~\subref*{fig:one_a} and the $W=0$ plot line in Fig.~\subref*{fig:one_b}  visualize the landscape of $\sigma_{xy}^{W=0}(E_\mathrm{F})$.

We would like to emphasize two points. First, the zero value of $\sigma_{xy}^{W=0}(E_\mathrm{F})$ in the band gap hallmarks the topologically trivial nature of the bands. Second, the approximate plateau of $\sigma_{xy}^{W=0}(E_\mathrm{F})$ within either band has a non-quantized (and non-universal with model parameters) value, reflecting the energetic separation of the aforementioned Berry fluxes within the bands. Since there is no spectral gap or mobility gap in the energy windows of finite Hall conductance in the clean case, the value of $\sigma_{xy}^{W=0}$ is not quantized and can be tuned with the hopping parameters. Clearly, for $E_\mathrm{F}$ inside the bands, the metallic nature is reflected in an extensive value of $\sigma_{xx}^{W=0}$. 

{\it Transport properties at finite disorder. ---}
Now, consider switching on disorder, i.e. $W>0$. In 2D, the theory of Anderson localization tells us that even a small random potential localizes (almost) all states and introduces a mobility gap, thus forcing the Hall conductance to a quantized value. At small $W$, it seems natural that $\sigma_{xy}$ may jump to the closest possible integer value, which is then retained on further increasing disorder strength. This intuition agrees with field-theoretical studies of disordered models with non-trivial Chern bands \cite{Pruisken_1, Pruisken_2, Pruisken_3, Khmelnitskii, Disordered_graphene, field_theory_disordered_CI}, which study plateau broadening finding \gls{iqh} transitions to concur with half-integer values of the accumulated Berry flux [cf. \eq{Eqn:accumulated_flux}]. Here, we extend this phenomenology to systems without Chern bands. Importantly, once protected by a mobility gap, a quantized value of $\sigma_{xy}$ is topologically stable against any small perturbation, just as for the conventional Quantum Hall scenario with an underlying Chern band structure or Landau level structure. Remarkably, extensive numerical analysis of our model system fully supports this rationale of a robust quantum Hall effect without Chern bands, as we detail in the following [see $W=1.5$ line and inset in Fig.~\subref*{fig:one_b} for a first impression]. 

Using the software package KWANT \cite{Kwant, Kwant_Mumps, KPM_tools}, we start our numerical simulations with a two-terminal linear response calculation of the longitudinal conductance $\sigma_{xx}$ as a function of Fermi Energy $E_\mathrm{F}$ and disorder strength $W$ for a mesoscopic system of $N_x = 600$, $N_y = 150$. We note that the aspect ratio of $N_x/N_y=4$ in this and the following simulations is chosen due to the anisotropic nature of our model. The leads are modeled with a width of $50$ sites attached to the center of each vertical face, i.e. at $x=1$ and $x=N_x$ around $y=N_y/2$. The lead dispersion is set to $t_\mathrm{L} [\cos(k_x)  + \cos(k_y)] \sigma_0$ with $t_\mathrm{L} = 10$. In Fig.~\ref{Fig:Transport}, we compare the transport properties for \gls{obc} and \gls{pbc} to highlight the effect of edge states. For \gls{obc}  [see Fig.~\subref*{Fig:Transport_a}], we find an extended region where the conductance is quantized to one, indicating a topologically protected edge state. By contrast, for \gls{pbc} [see Fig.~\subref*{Fig:Transport_b}], any edge states disappear and the sample is found to be insulating in the parameter regime where conductance quantization is most precise for \gls{obc}. We also note a strong $\sigma_{xx} > \frac{e^2}{h}$ at weak disorder $W<1$ for any $E_\mathrm{F}$ inside the bands. We attribute this to an expected finite size effect: when the average localization length becomes comparable to system size, an extensive number of evanescent modes contributes to the longitudinal conductance. The inset of Fig.~\subref*{Fig:Transport_a} quantifies the dependence of the conductance on $N_x$. All data in Fig.~\ref{Fig:Transport} is the average of $S = 20$ independent disorder realizations.

We note that Fig.~\subref*{Fig:Transport_b} also provides a complementary view on the non-quantized conductance of the clean system that is nudged into an \gls{iqh} phase through the disorder potential, relating to previous work on systems with topological bands \cite{Laughlin_IQH_2, Floquet_levitation, Fine_structure}. There, the finite Berry fluxes of opposite sign separate and flow towards mutual annihilation. Importantly, Berry charges do not annihilate within the same band but recombine inter-band, thus forming an inner arc and an outer arc in the $E_\mathrm{F}-W$ parameter plane. The phase of quantized Hall conductance is enclosed by these two arcs.

We now probe the transverse (or Hall) conductance $\sigma_{xy}$ at finite $W = 1.5$. This amounts to modeling our transport simulation as a four-terminal setting \cite{Datta, 4_terminal_Hall}. The system size is set to $N_x = 2000$, $N_y = 500$ and the leads are attached to the center of all four sample faces, with a lead width of 50 sites at the vertical faces and 100 sites at the horizontal faces with larger localization lengths. In Fig.~\subref*{fig:one_b}, numerical data showing the quantization of $\sigma_{xy}$ to $\frac{e^2}{h}$ in an extended window of $E_\mathrm{F}$ averaged over $S = 40$ independent disorder realizations is presented. As expected for a finite system with limited self-averaging properties, larger $S$ improves the precision of the quantization. However, we emphasize that already the single measurements clearly fluctuate around the quantized value as opposed to the non-universal value of the clean system \cite{Supplemental}. The inset of Fig.~\subref*{fig:one_b} shows how deviations from the quantized value decay with increasing system size for fixed $E_\mathrm{F}$ within the plateau region. We note that the quantized Hall plateau at $W = 1.5$ differs from the window of large non-quantized Berry flux at $W = 0$ in agreement with the continuous deformation of the region of quantized transport in Fig.~\ref{Fig:Transport} as a function of $W$. However, we emphasize that there is no bulk band gap closing between $W=0$ and $W=1.5$.

Complementary to the above four-terminal setup, we now explore a larger parameter range in the $W-E_\mathrm{F}$ plane by computing $\sigma_{xy}$ through the Kubo-Bastin formula \cite{Kubo_paper, Kubo_Bastin, Kubo_Bastin_modern} for linear response conductivity 
\begin{align}
\sigma_{xy}(E_\mathrm{F}, T) =& \frac{ie^2 \hbar}{V} \int_{-\infty}^\infty \mathrm{d} E f(E_\mathrm{F} - E, T)  \nonumber \\ &\times \mathrm{Tr}\left[v_x \frac{\mathrm{d} G^+}{\mathrm{d} E} v_y \rho(E) 
 - v_x  \rho(E) v_y \frac{\mathrm{d} G^-}{\mathrm{d} E}   \right].\label{Eqn:Kubo_Bastin}
\end{align}
Here, $f(E - E_\mathrm{F}, T)$ is the Fermi distribution at temperature $T$, $v_\alpha = -(i/ \hbar) [R_\alpha, H]$ denotes the velocity operator as induced by the position operator $R_\alpha$, $\rho(E) = \delta(E - H)$ is the \gls{dos}, and $G^{\pm} = [E \pm i 0^+ - H]^{-1}$ the retarded and advanced \gls{gf}, respectively. In the interest of numerical efficiency, the \gls{dos} and \gls{gf} may be approximated by an expansion in Chebyshev polynomials to order $M$, with the expansion coefficients determined from a Monte-Carlo evaluation of the trace using $R$ random phase vectors \cite{Kubo_Bastin_KPM_1, Kubo_Bastin_KPM_2, KPM_Review}. Together with $S$, these parameters control the numerical accuracy in evaluating \eq{Eqn:Kubo_Bastin} \cite{Supplemental}.

Fig.~\ref{Fig:Hall_cond} shows $\sigma_{xy}$ as a function of $W$ and $E_\mathrm{F}$ at $T = 0$. These results confirm the edge state picture discussed above in the context of the longitudinal conductance [cf.~Fig.~\subref*{Fig:Transport_a}]. We attribute minor deviations of $\sigma_{xy}$ from its quantized value in the core \gls{iqh} parameter regime to the combination of finite-size effects and the approximate nature of the numerical evaluation. For a quantitative analysis of finite-size effects, we refer the reader to the inset of Fig.~\subref*{fig:one_b}.

{\it Topology of conductance quantization. ---} 
When the Fermi energy lies inside a mobility gap at finite $W$, for which zero $\sigma_{xx}$ is a sufficient condition here, a topological invariant may be defined directly for the reflection part $R$ of the unitary $S$-matrix. Specifically, introducing twisted boundary conditions perpendicular to the direction of transport with a flux angle $\phi$, the integer quantized winding number 
\begin{align}
\nu = \frac{1}{2\pi} \oint_{0}^{2 \pi} \mathrm{d} \phi\, \left(\frac{\partial}{\partial\phi} \mathrm{arg}[\mathrm{det}[R(\phi, E)]] \right) \label{Eqn:W_num}
\end{align}
of the reflection matrix counts the charge pumped between the cylinder edges for a full twist of the boundary phase, which then corresponds to $\sigma_{xy}$ for \gls{pbc} in units of $\frac{e^2}{h}$ \cite{Smatrix_pumping, Smatrix_pumping_2, Scattering_invariants}. We calculate $\nu$ at multiple points in the phase diagram indicating the result in Fig.~\subref*{Fig:Transport_b}, which further corroborates the topological nature of the reported \gls{iqh} phase. 

As a complementary diagnostic tool for topology, we analyze the spectral localizer defined as the Hermitian matrix
\begin{align}
L(x,y,E) =&  \sigma_z \otimes (H - E) + \kappa \sigma_x \otimes (R_x - x) \nonumber \\
&+ \kappa \sigma_y \otimes (R_y - y)
\end{align}
for a system in two spatial dimensions and symmetry class A \cite{Budich2013, Symmetry_classes}. Here, $R_\alpha$ denotes the position operator for coordinate $\alpha$, and $x,y \in \mathbb R$ are real numbers representing a position in real space, while $\kappa > 0$ balances the relative weight of $H$ and $R_\alpha$ \cite{Spec_loc_1, Spec_loc_2, Spec_loc_3, Spec_loc_4, Spec_loc_5}. A prerequisite of topological robustness is then provided by the bulk localizer gap
\begin{align}
g_{L}(E) = \text{min}_{x,y \in \text{bulk}} \left| \text{spec}[L(x,y, E)]\right|, \label{Eqn:loc_gap}
\end{align}
where $\text{spec}[...]$ denotes the set of all eigenvalues. At energy $E$ located in a spectral gap of $H$, a finite localizer gap $g_{L}(E)$ can be proven to exist for suitable $\kappa$. Given a bulk localizer gap, the index 
\begin{align}
Q(x,y, E) = \frac{1}{2} \text{sig}[L(x,y, E)], \label{Eqn:loc_index}
\end{align}
where the signature $\text{sig}[L(x,y, E)]$ denotes the difference in the number of positive and negative eigenvalues, is well-defined and a constant integer for $x,y \in \text{bulk}$. Physically, $Q$ measures the number of chiral edge states encircling the system at energy $E$ \cite{Spec_loc_1, Spec_loc_2, Spec_loc_3, Spec_loc_4, Spec_loc_5}.
We emphasize that the assumption of a spectral gap does not apply to our present analysis at $W=0$. However, it has been observed that $g_{L}(E)$ can remain finite even if there is no spectral or mobility gap at $E$, which is still an active subject of research in math and physics \cite{Fine_structure, Spec_loc_4}.

In our model, choosing $\kappa \approx 0.4$, we find that a localizer gap opens deep inside the bulk band, which is accompanied by a non-trivial index $Q = -1$ (see Fig.~\ref{Fig:spec_loc}). Contributing to the ongoing discussion about the role of $g_{L}(E)$ in metallic situations, we argue that the data in Fig.~\ref{Fig:spec_loc} still predicts the formation of a stable \gls{iqh} phase in the presence of a generic random potential. In particular, at small disorder strength, for which all eigenvalues of $\hat W$ are smaller than $g_{L}(E)$, the localizer gap cannot close by virtue of Weyl's inequality \cite{Fine_structure, Spec_loc_5,Supplemental}. However, in the thermodynamic limit infinitesimal strength of disorder should suffice to cause a mobility gap due to Anderson localization, for which the localizer index predicts a chiral edge state surrounding the system \cite{spec_loc_mobility_gap}, which by bulk-boundary correspondence is tantamount to an \gls{iqh} phase.

Comparing the accumulated Berry flux with the lcoalizer gap (see Fig.~\ref{Fig:spec_loc}), we note that $\sigma_{xy}^{W=0}(E_\mathrm{F})<0.5\, e^2 / h$ corresponds well to regions of a topologically trivial ($Q=0$) localizer gap. However, a finite topological localizer gap ($Q= -1$) is only clearly visible in part of the energetic region with $\sigma_{xy}^{W=0}(E_\mathrm{F}) > 0.5\, e^2 / h$. While this may to some extent relate to finite size effects \cite{Supplemental}, we do not exclude that $Q$ contains additional structure beyond rounding $\sigma_{xy}^{W=0}$ to integer values. We emphasize that none of these scenarios contradicts our findings regarding $\sigma_{xy}$ at finite $W$.  

{\it Concluding Discussion. ---} 
We have demonstrated how a topologically stable quantum Hall effect may appear at the onset of disorder without non-trivial Chern bands in the underlying band structure. From a practical computational perspective, we emphasize that the minimal disorder strength $W > 0$ required to observe this effect is constrained only by the numerically accessible system sizes that must exceed the localization length, which diverges for $W \rightarrow 0$. On a more physical note, the subtle interplay between increasing localization length necessitating larger system sizes and the asymptotically discontinuous integer rounding of $\sigma_{xy}$ within trivial bands represents an intriguing order-of-limits problem worth future analytical study. There, calculating a disorder-averaged real-space Green’s function \cite{RS_GF_1, RS_GF_2} and analyzing its topological properties \cite{GF_topology_2} may provide complementary insight. To conclude, we would like to integrate our findings into a broader context by discussing their relation to previous work on non-standard (from a perspective of topological Bloch bands) topological phenomena.

In periodically driven (Floquet) systems far from equilibrium, a counterpart of the \gls{iqh} effect relating to Floquet chiral edge states has been reported despite zero Chern number of all bulk bands \cite{Anomalous_Floquet_AI, Anomalous_BBC}. However, there the underlying band structure is still topological and the chiral edge states can only be removed by changing a bulk topological invariant unique to Floquet systems, which makes the role of zero Chern numbers very different from our present work.

Moreover, in a topological band structure with a trivial band that is energetically sandwiched between two Chern bands and connected to them via topologically protected chiral edge states, a topological fine structure has recently been reported \cite{Fine_structure}. There, a bulk localizer gap inside the trivial band is connected to a disorder-driven transition where the two Chern bands do not recombine directly but instead with (substructure of) the trivial band. However, this leaves open the main question addressed in our present work, namely whether a quantum Hall effect can occur in a system without any topological bands and protected chiral edge states in the underlying band structure, respectively.

Finally, a phase known as the topological Anderson insulator (TAI) has been reported \cite{TAI, theory_TAI, Mobility_gap_TAI, Disordered_CI_HHH}.
There, starting from a trivial gapped band structure, disorder drives a topological quantum phase transition into a Chern insulator phase, amounting to a bulk gap closing and reopening in the effective band structure at finite $W$. While turning a trivial insulator into a topological one by disorder is a powerful and intriguing mechanism, qualitatively this scenario is similar to entering a Chern insulator phase via a bulk gap closing upon tuning a band structure parameter such as a Dirac mass. By contrast, our present finding is that a \gls{iqh} phase can form in a trivial band structure without a bulk gap closing being responsible. Instead, here the \gls{iqh} effect occurs with the onset of disorder inside the trivial bands, while a bulk gap closing of TAI type only happens at quite large disorder strength $W\approx 2.5$ (cf.~Fig.~\ref{Fig:Transport}).

Broadening the perspective, with the advent of Chern insulators generalizing from the specific setting of Landau levels, it was demonstrated that a magnetic field or net magnetization is not necessary for the occurrence of the quantum Hall effect \cite{Haldane_1988}. In this general context, our present work further relaxes the conditions for the generic occurrence of the quantum Hall effect by exemplifying how the energetic separation of substantial but non-quantized Berry fluxes in topologically trivial band structures can be sufficient. Our findings thus further widen the habitat of topologically quantized physical observables. \\

{\it Acknowledgments. ---}
We would like to thank Emil J. Bergholtz, Ion Cosma Fulga, Siddharth Parameswaran, and Bj\"orn Trauzettel for discussions. We acknowledge financial support from the German Research Foundation (DFG) through the Collaborative Research Centre SFB 1143  (Project-ID 247310070), the Cluster of Excellence ct.qmat (Project-ID 390858490). Our numerical calculations were performed on resources at the TU Dresden Center for Information Services and High Performance Computing (ZIH).\\

{\it Data availability. ---} 
The raw data for each data figure and scripts to generate all figures are available on Zenodo \cite{Zenodo}.

\onecolumngrid
\newpage

\vspace{15pt}
\section*{Supplemental Online Material: Quantum Hall Effect without Chern Bands}

\section{Details on the model}
The translation-invariant part of the model from the main text is given by 
\begin{align}
\hat H_0 = \sum_{\bm k} \bm c_{\bm k}^\dagger h_0(\bm k) \bm c_{\bm k}
\end{align}
with fermionic creation operators $\bm c_{\bm k}^\dagger = (c_{\bm k, a}^\dagger, c_{\bm k, b}^\dagger)$ in reciprocal space and the Bloch Hamiltonian 
\begin{align}
h_0(\bm k) =
\begin{pmatrix}
d_z(\bm k) & d_x(\bm k) - i d_y(\bm k) \\
d_x(\bm k) + i d_y(\bm k) & - d_z(\bm k)
\end{pmatrix},
\end{align}
where 
\begin{align}
d_x(\bm k) =& \gamma  \sin(k_x), \nonumber \\
d_y(\bm k) =& \lambda(k_x)  \sin(k_y), \nonumber \\
d_z(\bm k) =&  \gamma_2[r - \cos(2 k_x)] - \lambda(k_x) \cos(k_y),
\end{align}
and $\lambda(k_x) = \epsilon_1 + \epsilon_2 \left[1 - \cos(k_x)\right]/2$. Using $\cos(x)\cos(y) = [\cos(x+y) + \cos(x-y)]/2$ and $\cos(x)\sin(y) = [\sin(x+y) - \sin(x-y)]/2$, we may rewrite this as
\begin{align}
d_x(\bm k) =& \gamma \sin(k_x), \nonumber \\
d_y(\bm k) =& (\epsilon_1 + \epsilon_2 / 2) \sin(k_y) -  \epsilon_2 [\sin(k_x + k_y) - \sin(k_x - k_y)] / 4, \nonumber \\
d_z(\bm k) =& \gamma_2[r - \cos(2 k_x)] - (\epsilon_1 + \epsilon_2 / 2)\cos(k_y) + \epsilon_2 [\cos(k_x + k_y) + \cos(k_x - k_y)] / 4. 
\end{align}

\subsection{Real space Hamiltonian}
The formulation in real space following from the Fourier transform $c_{k, \alpha}^\dagger = \frac{1}{\sqrt{N_x N_y}} \sum_{\bm j} e^{i \bm k \cdot \bm j} c_{\bm j , \alpha}^\dagger$ and the relation 
\begin{align}
\sum_{\bm j } c^\dagger_{\bm{j}, \gamma} c_{\bm{j} + \bm \delta, \gamma'} =& \sum_{\bm k} e^{i \bm k \bm \delta} c^\dagger_{\bm k, \gamma} c_{\bm k, \gamma'},
\end{align}
as
\begin{align}
\hat H_{0} = & \sum_{\bm j} \left [\gamma_2 r \bm c_{\bm{j}}^\dagger \sigma^z \bm c_{\bm{j }} + \left \{\frac{\gamma}{2i} \bm c_{\bm{j}}^\dagger \sigma^x \bm c_{\bm{j + \delta_x}} + \frac{\epsilon_1 + \epsilon_2 / 2}{2i} \bm c_{\bm{j}}^\dagger \sigma^y \bm c_{\bm{j + \delta_y}}  - \frac{\epsilon_2}{8i} \bm c_{\bm{j}}^\dagger \sigma^y \bm c_{\bm{j + \delta_x + \delta_y}}  + \frac{\epsilon_2}{8i} \bm c_{\bm{j}}^\dagger \sigma^y \bm c_{\bm{j + \delta_x - \delta_y}} \right. \right. \nonumber \\
& - \left. \left. \frac{\epsilon_1 + \epsilon_2 / 2}{2} \bm c_{\bm{j}}^\dagger \sigma^z \bm c_{\bm{j  + \delta_y}} + \frac{\epsilon_2}{8} \bm c_{\bm{j}}^\dagger \sigma^z \bm c_{\bm{j  + \delta_x + \delta_y}} +  \frac{\epsilon_2}{8} \bm c_{\bm{j}}^\dagger \sigma^z \bm c_{\bm{j  + \delta_x - \delta_y}} - \frac{\gamma_2}{2} \bm c_{\bm{j}}^\dagger \sigma^z \bm c_{\bm{j  + 2 \delta_x}} + \text{H.c.}\right\} \right ].
\end{align}

\subsection{Berry curvature and energies}
The energies of the upper and lower band, which we denote here by $\pm$, are given by 
\begin{align}
E_\pm = \pm |\bm d| =& \sqrt{[\gamma \sin(k_x)]^2 + [\lambda(k_x) \sin(k_y)]^2 + [\gamma_2[r - \cos(2 k_x)] - \lambda(k_x) \cos(k_y)]^2}. \label{Eqn:Energy_analytically_App}
\end{align}
The Berry curvature for each band can be expressed as
\begin{align}
\mathcal F_{\pm}(\bm k) = \mp \bm {\hat  d} \cdot [\partial_{k_x} \bm {\hat  d} \times \partial_{k_y} \bm {\hat  d}]/ 2 = \frac{\mp 1}{|\bm d|^3}\bm {d} \cdot [\partial_{k_x} \bm {d} \times \partial_{k_y} \bm {d}]/ 2, \label{Eqn:Berry_curvature_analytically_App}
\end{align}
where $\bm {\hat  d} = \bm d / |\bm d|$ denotes the unit vector along $\bm d$. With
\begin{align}
\partial_{k_x} \bm d =& 
\begin{pmatrix}
\gamma \cos(k_x) \\
[\epsilon_2 / 2]\sin(k_x) \sin(k_y) \\
2 \gamma_2 \sin(2k_x) - [\epsilon_2 / 2]\sin(k_x) \cos(k_y) 
\end{pmatrix}, \quad 
\partial_{k_y} \bm d = 
\begin{pmatrix}
0 \\
\lambda(k_x)\cos(k_y) \\
\lambda(k_x) \sin(k_y)
\end{pmatrix},
\end{align}
we find 
\begin{align}
\partial_{k_x} \bm {\hat  d} \times \partial_{k_y} \bm {\hat  d} = 
\begin{pmatrix}
[\epsilon_2 / 2] \lambda(k_x) \sin(k_x)[\sin^2(k_y) + \cos^2(k_y)] - 2 \gamma_2 \lambda(k_x) \sin(2k_x) \cos(k_y) \\
- \gamma \lambda(k_x) \cos(k_x) \sin(k_y) \\
\gamma \lambda(k_x) \cos(k_x) \cos(k_y)
\end{pmatrix}
\end{align}
and ultimately
\begin{align}
\bm {d} \cdot [\partial_{k_x} \bm {d} \times \partial_{k_y} \bm {d}] =& \lambda(k_x) \{ [\epsilon_2 \gamma / 2] \sin^2(k_x) - 2 \gamma \gamma_2 \sin(k_x) \sin(2k_x) \cos(k_y)\nonumber  \\
&- \gamma \lambda(k_x) \cos(k_x) \sin^2(k_y) \nonumber\\
&+ \gamma \gamma_2 [r - \cos(2 k_x)] \cos(k_x) \cos(k_y) -  \gamma \lambda(k_x) \cos(k_x) \cos^2(k_y)\} \nonumber\\
=& \lambda(k_x) \{[\epsilon_2 \gamma / 2] \sin^2(k_x) - 2 \gamma \gamma_2 \sin(k_x) \sin(2k_x) \cos(k_y)  \nonumber \\
&+ \gamma \gamma_2 [r - \cos(2 k_x)] \cos(k_x) \cos(k_y) - \gamma \lambda(k_x) \cos(k_x) \}.
\end{align}
With this, we have an analytic expression for \eq{Eqn:Berry_curvature_analytically_App}, which we can be evaluated on a discrete lattice in the First Brillouin zone together with the energy \eq{Eqn:Energy_analytically_App}. To obtain the $\sigma_{xy}^{W=0}(E_\mathrm{F})$ curve in the first figure of the main text in a computationally efficient manner, the resulting array can simply be sorted by energy and then summed up.

\subsection{Non-topological edge states in cylinder geometry}
We find that the separation of topological charges within the bands can cause a pair of edges states to emerge at a boundary. However, this also depends on the direction of the boundary. Importantly, if an edge state emerges, it will always have an anti-chiral partner due to the vanishing total Chern number of the bands, i.e. it is not topologically protected and can be gapped out continuously. 

To illustrate this, we compute the spectrum with \gls{pbc} in $x$-direction and \gls{obc} in $y$-direction in Fig.~\subref*{Fig:Cylinder_spec_a}. The edge states are highlighted by colors. The condition to be considered an edge state at the left / right edge is that $90\%$ of the wave function amplitude is concentrated within the left / right $25\%$ of the system. By contrast, in the spectrum with 
\gls{pbc} in $y$-direction and \gls{obc} in $x$-direction presented in 
Fig.~\subref*{Fig:Cylinder_spec_b}, no such states are present.

\begin{figure}[htp]	 
{
    \vbox to 0pt {
            \raggedright
            \textcolor{white}{
                \subfloatlabel[1][Fig:Cylinder_spec_a]
                \subfloatlabel[2][Fig:Cylinder_spec_b]
            }
        }
}
{\includegraphics[trim={0.75cm 0cm 0.75cm 0.cm}, width=0.80\linewidth]{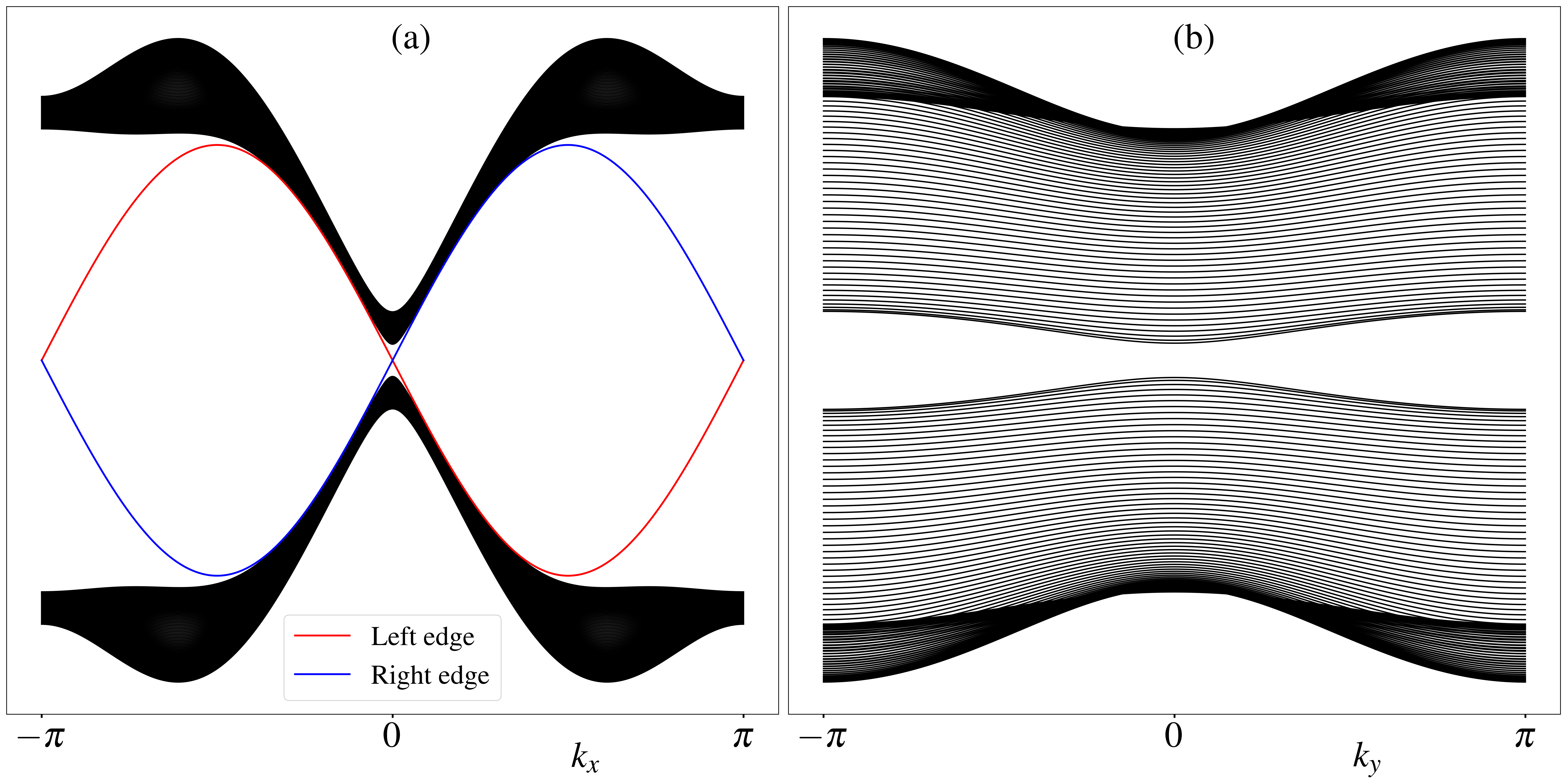}}
\caption{Spectra for cylinders with a height of 100 unit cells for parameters $r = 1.5$, $\epsilon_1 = 0.3$, $\epsilon_2 = 2$, $\gamma  =2$, and $\gamma_2 = 0.3$ as a function of momentum in the circumferential direction. Edge states indicated if present. a) Result for \gls{obc} in $y$-direction, a pair of edge states emerges. b) Result for \gls{obc} in $x$-direction, no edge states present.
 }\label{Fig:Edge_states}
\end{figure}

\section{Additional data on the four terminal Hall conductance}
\begin{figure}[htp]	 
\includegraphics[trim={0.75cm 0cm 0.75cm 0.cm}, width=0.95\linewidth]{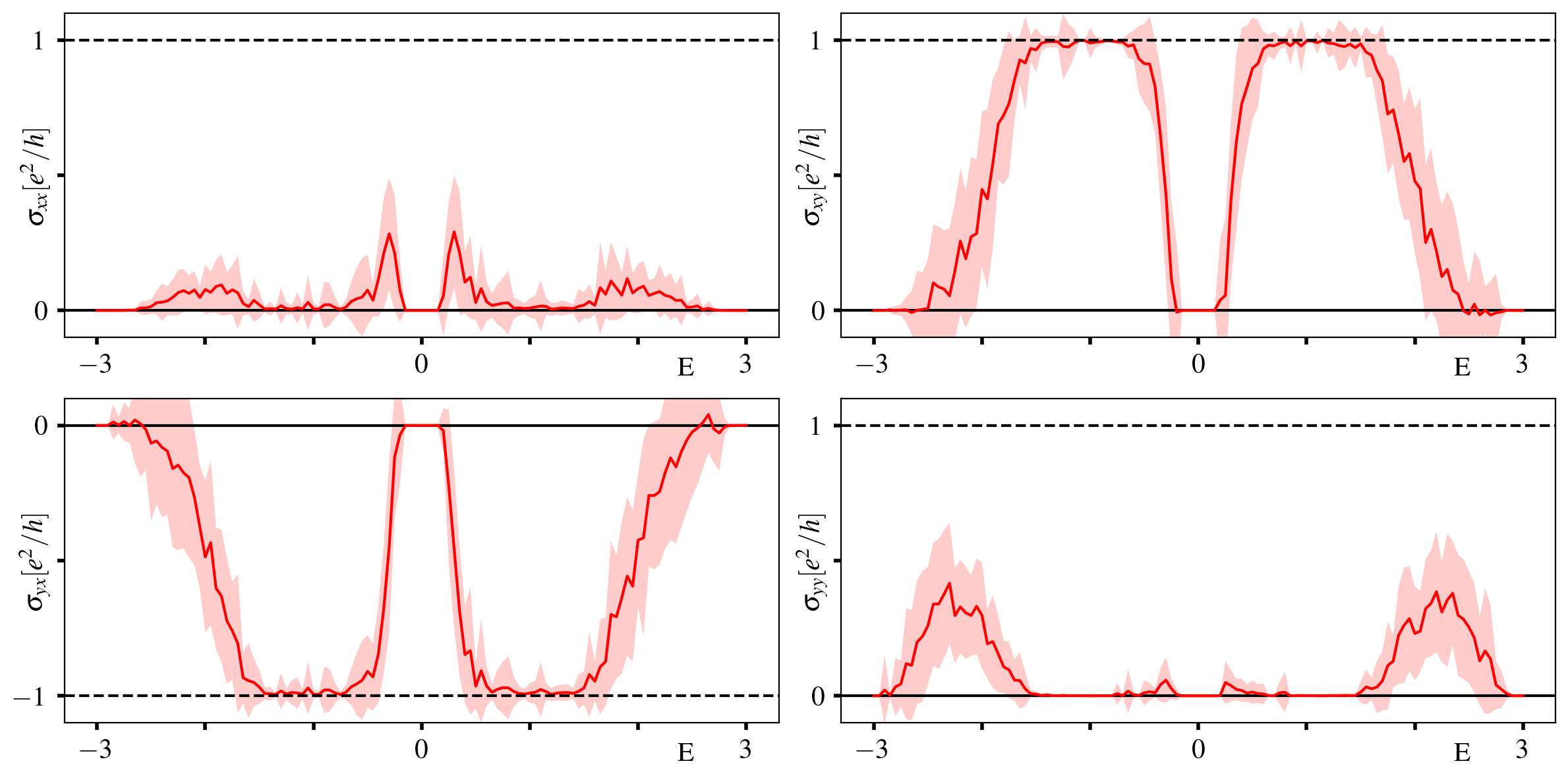}
\caption{Elements of the conductance tensor for a system size of $N_x = 2000$, $N_y = 500$ and parameters $r = 1.5$, $\epsilon_1 = 0.3$, $\epsilon_2 = 2$, $\gamma  =2$, $\gamma_2 = 0.3$ with disorder strength $W = 1.5$ as obtained from a standard four terminal scattering calculation. The leads attached to the vertical faces have a width of $50$ unit cells and those at the horizontal faces have a width of $200$ unit cells. The result is the average of $S = 40$ disorder realizations, standard deviation is indicated as a corridor.
 }\label{Fig:Hall_cond_4_terminal}
\end{figure}

To obtain the curve for the Hall conductance $\sigma_{xy}$ presented in Fig.~\ref{fig:one} of the main text, we model the system as a four terminal scattering experiment with KWANT \cite{Kwant_App, Kwant_Mumps_App}. Following Refs.~ \cite{Datta_App, 4_terminal_Hall_App}, we extract the elements of the conductance tensor from the scattering amplitudes as 
\begin{align}
\sigma_{xx}  = \frac{e^2}{h} T_\mathrm{R \leftarrow L}, \quad \sigma_{xy}  = \frac{e^2}{h} (T_\mathrm{L \leftarrow T} - T_\mathrm{L \leftarrow B}), \quad \sigma_{yx}  = \frac{e^2}{h} (T_\mathrm{B \leftarrow R} - T_\mathrm{B \leftarrow L}), \quad \sigma_{yy}  = \frac{e^2}{h} T_\mathrm{T \leftarrow B}.
\end{align}
Here, $T_{p \leftarrow q}$ with $p, q = $ L (left), R (right), B (bottom), T (top) denotes the transmission amplitude from terminal $q$ to $p$. In Fig.~\ref{Fig:Hall_cond_4_terminal}, we present the result for all elements of the conductance tensor. The quantized plateaus of the off-diagonal elements are accompanied by a vanishing of $\sigma_{xx}$ and $\sigma_{yy}$. The peaks of the diagonal entries in the transition region between the Hall plateaus are indicative of a quantum phase transition. We illustrate the standard deviation of as a corridor (in Fig.~\ref{fig:one} of the main text, we use the standard error which is smaller by a factor of $1 / \sqrt{S}$ for visual clarity).

We also present a complementary way of extracting the Hall conductance from the four terminal scattering data. If all the transmission amplitudes $T_{p \leftarrow q}$ are known, the conductance matrix can be obtained as 
\begin{align}
G_{p, q} =  - \frac{e^2}{h} T_{p \leftarrow q} + \delta_{p, q} \frac{e^2}{h} \sum_{q'} T_{q' \leftarrow p}.
\end{align}
The units and sign conventions are chosen such that the current at terminal $p$ is obtained as $I_p = \sum_{q} G_{p,q} V_q$, where $V_q = -e \mu_q$ is the Voltage at terminal $q$, obtained from the chemical potential $\mu_q$ and the elementary charge $e = 1.602176634 \text{\textsc{e-}}19\mathrm{C}$.  

\begin{figure}[htp]	 
\includegraphics[trim={0.75cm 0cm 0.75cm 0.cm}, width=0.95\linewidth]{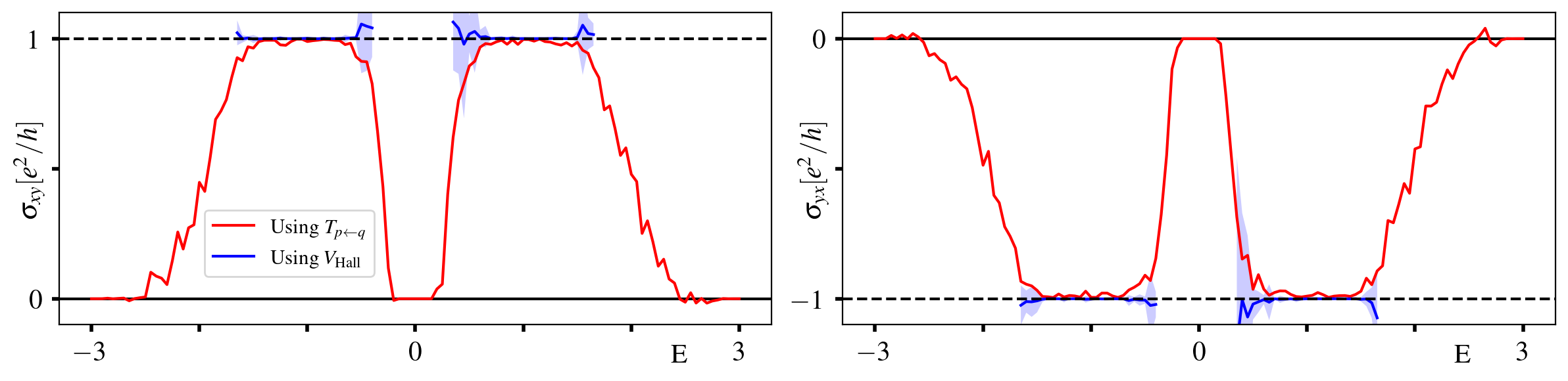}
\caption{Result for $\sigma_{xy}$ and $\sigma_{yx}$ for a system size of $N_x = 2000$, $N_y = 500$ and parameters $r = 1.5$, $\epsilon_1 = 0.3$, $\epsilon_2 = 2$, $\gamma  =2$, $\gamma_2 = 0.3$ with disorder strength $W = 1.5$. The leads attached to the vertical faces have a width of $50$ unit cells and those at the horizontal faces have a width of $200$ unit cells.  We present the result obtained from extracting the Hall voltage from the conductance matrix in blue and compare it to the result obtained from the difference of the scattering amplitudes in red (same data as Fig.~\ref{Fig:Hall_cond_4_terminal}). The standard deviation of the blue curve is indicated as a corridor.
 }\label{Fig:V_Hall_4_terminal}
\end{figure}

As a complementary approach, given the conductance matrix it is possible to impose a current in one direction and determine the transverse voltage by solving the linear equation $\bm I = G \bm V$. For $\sigma_{xy}$, this amounts to setting $I_L = I_x$, $I_R = -I_x$, $I_B = 0$, $I_T = 0$ with some arbitrary value $I_x$. This leaves the equation $\bm I = G \bm V$ underdetermined, since the current does not change under a constant shift of all voltages. We can thus set, e.g., $V_T = 0$, which amounts to truncating the conductance matrix to the top left $3\times3$ block and solving for $V_L$, $V_R$, and $V_B$. The Hall voltage is then $V_\mathrm{Hall} = V_B - V_T = V_B$ and the Hall conductance is found to be $\sigma_{xy} = I_x / V_\mathrm{Hall} = I_x / V_B$. To determine $\sigma_{yx}$, the same the same procedure is used for the currents $I_L = 0$, $I_R = 0$, $I_B = I_y$, $I_T = - I_y$. This approach only yields a definite result for $E_\mathrm{F} $ inside the conductance plateaus, where the bulk of the system is insulating. There, however, it resolves the Hall plateaus very sharply and with small variance as we show in Fig.~\ref{Fig:V_Hall_4_terminal}.

\section{Kernel polynomial approach to the Kubo Bastin formula} 
In the interest of a self-contained presentation, we briefly review the \gls{kpm} approach to computing the Kubo-Bastin formula that is implemented in the KWANT package, for more details see, e.g., Refs.~\cite{Kubo_Bastin_KPM_1_App, Kubo_Bastin_KPM_2_App, KPM_Review_App}.

\subsection{Chebyshev polynomial expansion}
The Chebyshev polynomials of the first kind are defined as
\begin{align}
T_m(x) = \cos [m \arccos(x)]
\end{align}
on the interval $[-1, 1]$ and obey the orthogonality relation
\begin{align}
\langle T_n , T_m \rangle_w = \frac{1 + \delta_{n,0}}{2} \delta_{n,m}
\end{align}
with respect to the scalar product defined by the weighting function $w(x) = [\pi \sqrt{1-x^2}]^{-1}$
\begin{align}
\langle f , g \rangle_w = \int_{-1}^1 f(x) g(x) w(x) dx.
\end{align}
Another useful property is the recursion relation
\begin{align}
T_{n+1}(x) = 2x T_n(x) - T_{n-1}(x). \label{Eqn:T_n_recursion}
\end{align}
For sufficiently regular functions on $[-1,1]$, an expansion in $T_n$ should converge uniformly and read
\begin{align}
f(x) = 2 \sum_{n = 0}^\infty \frac{\alpha_n T_n(x)}{\delta_{n,0} + 1}
\end{align}
with coefficients
\begin{align}
\alpha_n = \langle f, T_n \rangle_w =  \int_{-1}^1  f(x) T_n(x) w(x) dx. \label{Eqn:KPM_coefficients}
\end{align}
To obtain an approximation to $f(x)$, this expansion can be truncated at some order $M$
\begin{align}
f_M(x) = 2 \sum_{n = 0}^{M-1}  \frac{\alpha_n T_n(x)}{\delta_{n,0} + 1},
\end{align}
however naively doing so will usually lead to so-called Gibbs oscillations. To obtain a better approximation, it is possible to fold the truncated function with a Kernel
\begin{align}
f_{\text{KPM},M}(x) = \int_{-1}^1 \pi \sqrt{1 - y^2} f_M(y) K_M(x, y) dy, 
\end{align} 
where the Kernel $K_M(x, y)$ depends on the order $M$ of the approximation and can be chosen such that $f_{\text{KPM},M}(x)$ converges to $f(x)$ faster than the raw truncation $f_M(x)$. In practice, this amounts to multiplying the expansion coefficients by factors $g_n(M)$ that depend on the order M of the approximation:
\begin{align}
f_{\text{KPM},M}(x) = 2  \sum_{n = 0}^{M-1} g_n(M) \frac{\alpha_n T_n(x)}{\delta_{n,0} + 1}. \label{Eqn:KPM_expansion}
\end{align} 

The optimal choice of Kernel depends on the nature of the problem at hand and which properties of the approximated functions should be preserved to obtain an accurate solution. The calculations in this paper employ the Jackson Kernel, which is well-suited for most physics applications and amounts to the choice
\begin{align}
g_n = \frac{(M-n+1) \cos[\pi n / (M+1)] + \sin[\pi n / (M+1)] \cot[\pi / (M+1)]}{M+1}.
\end{align} 

\subsubsection{Extension to operator-valued functions}
To apply the method to operator-valued functions of the Hamiltonian $H$, it has to be rescaled such that its spectrum is contained within $[-1,1]$. This can be done by
\begin{align}
H \to \tilde H = \frac{2}{\Delta_E}\left(H - \frac{E^+ + E^-}{2} \right), \quad E \to \tilde E = \frac{2}{\Delta_E}\left(E - \frac{E^+ + E^-}{2} \right), \label{Eqn:rescaling_KPM}
\end{align}
where $\Delta_E = E^+ - E^-$ and $E^\pm$ denote upper / lower bounds on the spectrum that can be estimated by Krylow methods. Then, any function $f(\tilde E, \tilde H)$ of the rescaled Hamiltonian Hamiltonian and an energy $\tilde E$ can be expanded as
\begin{align}
f(\tilde  E,\tilde  H) =& \sum_l f(\tilde  E, \tilde  E_l) \ket{\tilde  E_l}\bra{\tilde  E_l} \nonumber \\
=& \sum_l \left[2 \sum_{n = 0}^{M-1}g_n(M) \alpha_n(\tilde E) T_n(\tilde E_l) \right ]\ket{\tilde E_l}\bra{\tilde E_l} \nonumber  \\
=& 2 \sum_{n = 0}^{M-1}\frac{g_n(M) \alpha_n(\tilde E) }{\delta_{n,0} + 1} T_n(\tilde H), \label{Eqn:KPM_expansion_operator}
\end{align} 
where we applied the \gls{kpm} expansion \eq{Eqn:KPM_expansion} to the second argument of $f(\tilde E,\tilde E_l)$. Thus, the coefficients read (cf. \eq{Eqn:KPM_coefficients})
\begin{align}
\alpha_n(\tilde E) = \int_{-1}^{1} \frac{f(\tilde E, x) T_n(x)}{\pi \sqrt{1 -x^2}} dx. \label{Eqn:KPM_coefficients_operator}
\end{align} 

\subsection{Expansion of the Kubo-Bastin formula}
The exact linear response expression for the conductance tensor in a non-interacting system can be derived from the Kubo formula as 
\begin{align}
\sigma_{\alpha \beta}(E_\mathrm{F}, T) =& \frac{i e^2 \hbar}{V} \int_{E^-}^{E^+} \mathrm{d} E f_\mathrm{F}(E - E_\mathrm{F}, T) \mathrm{Tr} \left[v_\alpha \frac{\mathrm{d} G^+(E)}{\mathrm{d} E} v_\beta \rho(E)  - v_\alpha \rho(E) v_\beta  \frac{\mathrm{d} G^-(E)}{\mathrm{d} E} \right],\label{Eqn_App:Kubo_Bastin}
\end{align}
where $f_\mathrm{F}(E - E_\mathrm{F}, T) = 1 / (\exp[(E-E_\mathrm{F}) / (k_\mathrm{B} T)] + 1)$ \cite{Kubo_paper_App, Kubo_Bastin_App, Kubo_Bastin_modern_App}. Given the position operator $R_\alpha$, the velocity operators follow from the Heisenberg picture as $v_\alpha = \dot R_\alpha = -(i/ \hbar) [R_\alpha, H]$.

The other two relevant quantities are the \gls{dos} $\rho(E, H) = \delta(E - H)$ and the retarded/adavanced \gls{gf} $G^{\pm}(E, H) = [E \pm i 0^+ - H]^{-1}$. First of all, the above expression has to be rescaled in the sense of \eq{Eqn:rescaling_KPM}. To this end, note that $\delta((\tilde E + b) / a - H) = |a| \delta (\tilde E - (a H - b)) = a \delta (\tilde E - \tilde H)$ and $G^\pm[(\tilde E + b) / a, H] = [(\tilde E + b) / a - H + i 0^\pm]^{-1} = a  [\tilde E - \tilde H + i a 0^\pm]^{-1} = a G^\pm[\tilde E , \tilde H]$, where $a = 2/\Delta_E$, $b = (E^+ + E^-)/ \Delta_E$, from which we can infer by straightforward substitution
\begin{align}
\sigma_{\alpha \beta}(E_\mathrm{F}, T) =& \frac{i a^2 e^2 \hbar}{V} \int_{-1}^{1} \mathrm{d} \tilde E f_\mathrm{F}((\tilde E + b)/a - E_\mathrm{F}, T) \mathrm{Tr}\left[v_\alpha \frac{\mathrm{d} G^+(\tilde E)}{\mathrm{d} \tilde  E} v_\beta \rho(\tilde E) - v_\alpha \rho(\tilde  E) v_\beta \frac{\mathrm{d} G^- (\tilde E)}{\mathrm{d} \tilde E}   \right]. \label{Eqn_App:Kubo_Bastin_rescaled}
\end{align}
Now, the \gls{dos} and \gls{gf} can be expanded as per \eq{Eqn:KPM_expansion_operator}, with the coefficients following from \eq{Eqn:KPM_coefficients_operator}: 
\begin{align}
\alpha_{n, \rho}(\tilde E) =& \int_{-1}^{1} \frac{\delta(\tilde E-x) T_n(x)}{\pi \sqrt{1 -x^2}} dx = \frac{T_n(\tilde E)}{\pi \sqrt{1 - \tilde E^2}} \\
\alpha_{n, G^{\pm}}(\tilde E) =& \int_{-1}^{1} \frac{T_n(x)}{[\tilde E + i 0^\pm - x]\pi \sqrt{1 -x^2}} dx = \mathcal P \left[\int_{-1}^{1} \frac{T_n(x)}{\pi [\tilde E - x] \sqrt{1 -x^2}} dx\right]  \mp i \frac{T_n(\tilde E)}{\sqrt{1 - \tilde E^2}} \nonumber \\
=& U_{n-1}(\tilde E) \mp i \frac{T_n(\tilde E)}{\sqrt{1 - \tilde E^2}} = \mp i \frac{e^{\pm i n \arccos(\tilde E)}}{\sqrt{1 - \tilde E^2}}.
\end{align}
The real part of $\alpha_{n, G^{\pm}}(\tilde E)$ is a standard integral relation between the Chebyshev polynomials of the first kind, $T_n(x)$, and the second kind $U_n(x) = \sin[(n+1) \arccos(x)] / \sqrt{1 - x^2}$ \cite{KPM_Review_App}. The derivative of $\alpha_{n, G^{\pm}}(\tilde E)$ \gls{wrt} $\tilde E$ is
\begin{align}
\frac{d}{d \tilde E} \alpha_{n, G^{\pm}}(\tilde E) =& \mp i \left[ \frac{\tilde E}{(1 - \tilde E^2)^{3/2}} \mp \frac{i n}{1 - \tilde E^2}\right] e^{\pm i n \arccos(\tilde E)}.
\end{align}
Having obtained the coefficients, we can expand \eq{Eqn_App:Kubo_Bastin_rescaled} as
\begin{align}
\sigma_{\alpha \beta}(E_\mathrm{F}, T) =& \frac{4 a^2 e^2 \hbar}{\pi V} \int_{-1}^{1} \mathrm{d} \tilde  E \frac{f_\mathrm{F}((\tilde E + b)/a - E_\mathrm{F}, T) }{(1 - \tilde E^2)^2} \sum_{m,n = 0}^M \Gamma_{n,m}(\tilde E) \mu_{n,m}^{\alpha, \beta} , \label{Eqn_App:Kubo_Bastin_KPM}
\end{align}
where 
\begin{align}
\Gamma_{n,m}(\tilde E) = \left[\tilde E - i n \sqrt{1- \tilde E^2} \right] e^{i n \arccos(\tilde E)} T_{m} (\tilde E) + T_{n} (\tilde E) \left[\tilde E + i m \sqrt{1- \tilde E^2} \right] e^{- i m \arccos(\tilde E)} 
\end{align}
does not depend on the details of the system, which are contained in 
\begin{align}
\mu_{n, m}^{\alpha, \beta} = \frac{g_n(M) g_m(M)}{(1 + \delta_{n,0}) (1 + \delta_{m,0})} \mathrm{Tr}\left[v_\alpha T_{n}(H) v_\beta T_{m}(H)\right].
\end{align}
It is easily seen that $\Gamma_{m,n}^* = \Gamma_{n,m}$ and $(\mu_{m,n}^{\alpha, \beta})^* = \mu_{n,m}^{\alpha, \beta}$, which implies that \eq{Eqn_App:Kubo_Bastin_KPM} is a real number. The numerical cost is hidden in the evaluation of $\mu_{m,n}^{\alpha, \beta}$, however the trace of a generic operator $A$ can be sampled efficiently by using $R$ random phase vectors $\ket{\phi_l}$:
\begin{align}
\mathrm{Tr} [A] \approx \frac{1}{R} \sum_{l = 1}^R \bra{\phi_l} A \ket{\phi_l}.
\end{align}
This estimate improves with system size $N$, it can be shown that for a generic operator $A$ (in the sense of local and homogeneous support such that $\mathrm{Tr}[A^2] = O(N^2)$) the relative error should scale as $\sim \frac{1}{\sqrt{N R}}$  \cite{KPM_Review_App}. Furthermore, the recursion relation \eq{Eqn:T_n_recursion} can be used to evaluate the scalar products appearing in the approximation of the trace.

To obtain the data presented in the main text of the paper, we used the efficient implementation of the algorithm outlined here in the KWANT package \cite{Kwant_App, Kwant_Mumps_App}. To generate correct velocity operators for \gls{pbc}, we also acknowledge use of the extension KPM tools \cite{KPM_tools_App}.

\section{Hall conductance of the clean system from the Kubo-Bastin formula}
For completeness, we derive here that the Hall conductance of the clean system is indeed proportional to the accumulated Berry flux up to the Fermi energy $E_\mathrm{F}$. At $T = 0$, the Hall conductance according to the Kubo-Bastin Formula can be written as
\begin{align}
\sigma_{xy}(E_\mathrm{F}, T = 0) =& \frac{i e^2 \hbar}{V} \int_{- \infty}^{E_\mathrm{F}} \mathrm{d} E \mathrm{Tr} \left[v_x \frac{\mathrm{d} G^+(E)}{\mathrm{d} E} v_y \rho(E) - v_x \rho(E) v_y  \frac{\mathrm{d} G^-(E)}{\mathrm{d} E}  \right],
\end{align}
where all operators are to be interpreted as those arising from the single-particle tight-binding matrix \cite{Kubo_paper_App, Kubo_Bastin_App, Kubo_Bastin_modern_App}. In the clean, translation-invariant system, all appearing quantities are block-diagonal in the momentum basis. The blocks of the velocity operators are readily derived to be $v_\alpha(\mathrm k) = \partial_{k_\alpha} H(\bm k) / \hbar$,
the \gls{gf} terms become
\begin{align}
\frac{\mathrm{d} G^\pm(E, \bm k)}{\mathrm{d} E} = \frac{\mathrm{d} }{\mathrm{d} E} \sum_{l = 1}^{N_\mathrm{O}} \frac{1}{E + i 0^\pm - E_l(\bm k)} \ket{E_l(\bm k)} \bra{E_l(\bm k)} = - \sum_{l = 1}^{N_\mathrm{O}} \frac{1}{(E + i 0^\pm - E_l(\bm k))^2} \ket{E_l(\bm k)} \bra{E_l(\bm k)}, 
\end{align}
and the \gls{dos} 
\begin{align}
\rho(E, \bm k) = \sum_{l = 1}^{N_\mathrm{O}} \delta(E - E_l(\bm k)) \ket{E_l(\bm k)} \bra{E_l(\bm k)}.
\end{align}
The trace can be reduced to a sum over all $\bm k$ and a trace over the eigenstates $\ket{E_l(\bm k)}$ of the Bloch Hamiltonian at each $\bm k$ (even though there might be topological obstructions to finding a smooth gauge for the $\ket{E_l(\bm k)}$, this not a problem when taking the trace, as it is basis-independent). The Hall conductance thus becomes
\begin{align}
\sigma_{xy}(E_\mathrm{F}, T = 0) =& \frac{i e^2 }{V \hbar} \int_{- \infty}^{E_\mathrm{F}} \mathrm{d} E \sum_{\bm k} \sum_{l, l' = 1}^{N_\mathrm{O}} \left [ \bra{E_l(\bm k)}\left(\partial_{k_x} H(\bm k) \right)\ket{E_{l'}(\bm k)} \bra{E_{l'}(\bm k)} \left(\partial_{k_y} H(\bm k) \right)\ket{E_{l}(\bm k)}  \frac{- \delta(E - E_{l}(\bm k)) }{(E + i 0^+ - E_{l'}(\bm k))^2} \right .  \nonumber \\
& - \left.  \bra{E_l(\bm k)}\left(\partial_{k_x} H(\bm k) \right)\ket{E_{l'}(\bm k)} \bra{E_{l'}(\bm k)} \left(\partial_{k_y} H(\bm k) \right)\ket{E_{l}(\bm k)} \frac{- \delta(E - E_{l'}(\bm k)) }{(E + i 0^- - E_{l}(\bm k))^2} \right ].
\end{align}
The terms of the above equation where $l = l'$ vanish, for the rest we can neglect the $i 0^\pm$ regularization and carry out the integral to find
\begin{align}
\sigma_{xy}(E_\mathrm{F}, T = 0) =& -\frac{i e^2}{V \hbar} \sum_{\bm k} \sum_{l = 1}^{N_\mathrm{O}}\Theta(E_\mathrm{F} - E_l(\bm k))  \sum_{l' \neq l} \left [ \bra{E_l(\bm k)}\left(\partial_{k_x} H(\bm k) \right)\ket{E_{l'}(\bm k)} \bra{E_{l'}(\bm k)} \left(\partial_{k_y} H(\bm k) \right)\ket{E_{l}(\bm k)} \right .  \nonumber \\
& - \left.  \bra{E_l(\bm k)}\left(\partial_{k_y} H(\bm k) \right)\ket{E_{l'}(\bm k)} \bra{E_{l'}(\bm k)} \left(\partial_{k_x} H(\bm k) \right)\ket{E_{l}(\bm k)}  \right]  \frac{1}{(E_l(\bm k)  - E_{l'}(\bm k))^2} \nonumber \\
=& -\frac{ie^2}{V \hbar} \sum_{\bm k} \sum_{l = 1}^{N_\mathrm{O}}\Theta(E_\mathrm{F} - E_l(\bm k))  \sum_{l' \neq l} \left [ \braket{\partial_{k_x}  E_l(\bm k)|E_{l'}(\bm k)} \braket{E_{l'}(\bm k) |\partial_{k_y} E_{l}(\bm k)} \right . \nonumber \\
& - \left.  \braket{\partial_{k_y}  E_l(\bm k)|E_{l'}(\bm k)} \braket{E_{l'}(\bm k) |\partial_{k_x} E_{l}(\bm k)} \right] \nonumber \\
=& \frac{e^2}{V \hbar} \sum_{\bm k} \sum_{l = 1}^{N_\mathrm{O}}\Theta(E_\mathrm{F} - E_l(\bm k))   \left [-i (\braket{\partial_{k_x}  E_l(\bm k)|\partial_{k_y} E_{l}(\bm k)} - \braket{\partial_{k_y} E_l(\bm k)|\partial_{k_y} E_{l}(\bm k)}) \right ] \nonumber \\
=& \frac{e^2}{h} \sum_{l = 1}^{N_\mathrm{O}} \frac{1}{2 \pi} \int \mathrm{d}^2k  \Theta(E_\mathrm{F} - E_l(\bm k)) \mathcal F_l (\bm k) = \frac{e^2}{h} \sum_{l = 1}^{N_\mathrm{O}} \frac{1}{2 \pi} \int_{E_l(\bm k)<E_\mathrm{F}}\mathrm{d}^2k   \mathcal F_l (\bm k). \label{Eqn:Kubo_Bastin_clean}
\end{align}
For the second equality above, we used that $\bra{E_l(\bm k)}\left(\partial_{k_\alpha} H(\bm k) \right)\ket{E_{l'}(\bm k)} = \braket{\partial_{k_\alpha}  E_l(\bm k)|E_{l'}(\bm k)} [E_l(\bm k) - E_{l'}(\bm k)] = - \braket{E_l(\bm k)|\partial_{k_\alpha} E_{l'}(\bm k)} [E_l(\bm k) - E_{l'}(\bm k)]$ and for the third equation, we note that the term $l' = l$ yields zero and can formally be added to the sum $\sum_{l' \neq l}$ to yield a resolution of identity. Finally, the continuum limit is taken as $\Delta_{\bm k}^2 \sum{\bm k} = \int d^2 k$ with $\Delta_{\bm k}^2 = 4 \pi^2 / V$. 

\section{Additional data on the spectral localizer}
\begin{figure}[htp]	 
\includegraphics[trim={0.75cm 0cm 0.75cm 0.cm}, width=0.80\linewidth]{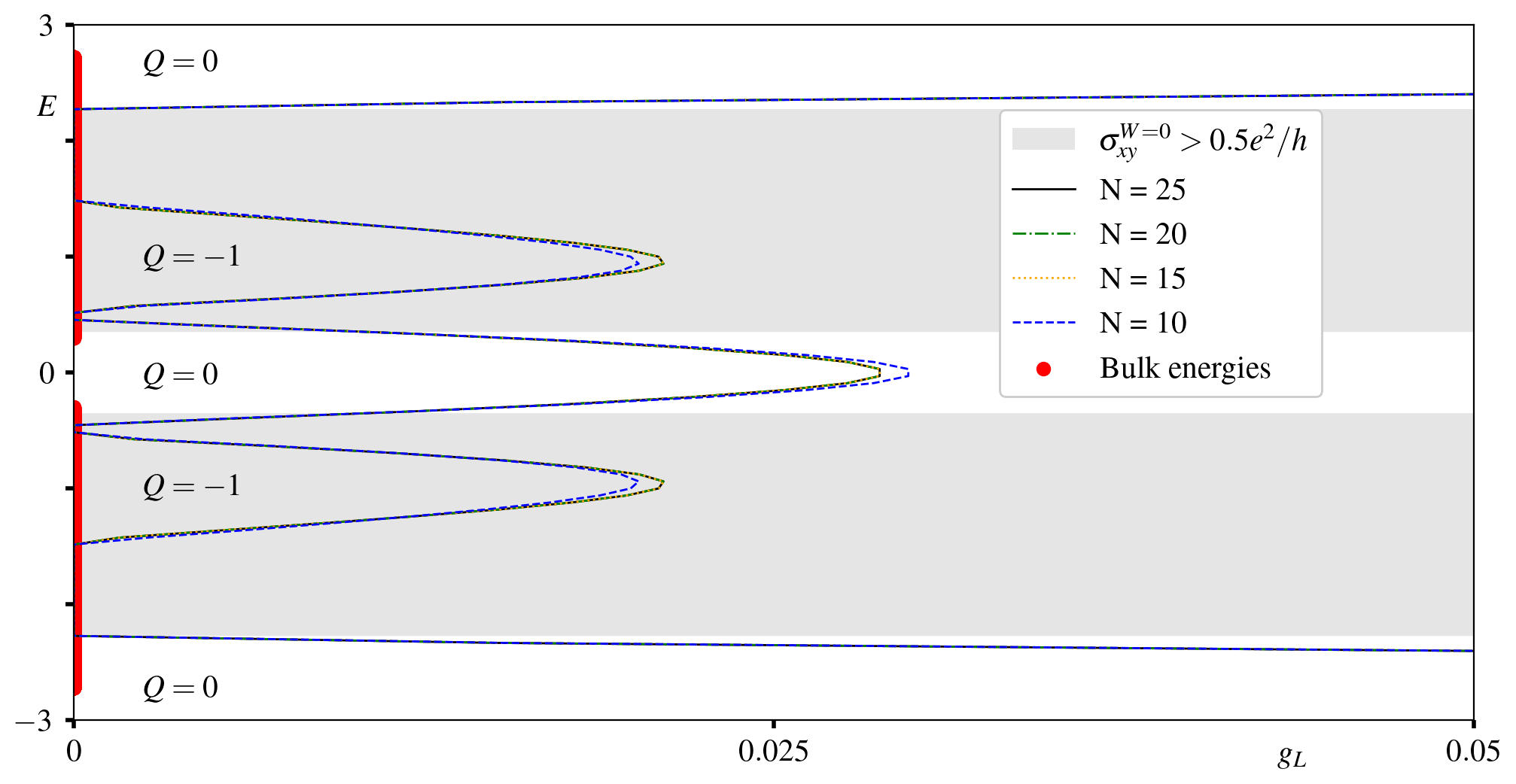}
\caption{Localizer gap $g_L$ [cf. \eq{Eqn:loc_gap} of main text] as a function of energy for parameters $r = 1.5$, $\epsilon_1 = 0.3$, $\epsilon_2 = 2$, $\gamma  =2$, and $\gamma_2 = 0.3$ for multiple 
system sizes of $N_x = N_y = N$. The scaling parameter is set to $\kappa = 0.4$. The value of $g_L$ is taken as the minimum of 100 reference positions inside the central Wigner-Seitz cell. Wherever the localizer gap is stable, we indicate the associated value of the topological index $Q$ [cf. \eq{Eqn:loc_index} of main text]. For reference, the clean bulk spectrum is shown in red and the energy window where $\sigma_{xy}^{W=0}(E_\mathrm{F})  > 0.5 e^2 / h$ [cf. \eq{Eqn:accumulated_flux} of main text] is marked as a gray corridor. }\label{Fig:Spec_Loc_Data_size}
\end{figure}

The spectral localizer serves as a type of local Chern marker and yields a topological index $Q$ [cf. \eq{Eqn:loc_index} of main text] as a function of spatial position and (Fermi) energy that remains valid in the presence of disorder. By construction, $Q$ is only well-defined in the presence of a localizer gap $g_L$ at the chosen energy [cf. \eq{Eqn:loc_gap} of main text], which is only guaranteed to emerge in the presence of a spectral gap \cite{Spec_loc_5_App} or a mobility gap \cite{spec_loc_mobility_gap_App}. The physical significance of the topological index $Q$ is to predict the quantized value of the Hall conductance iff there is a spectral gap or a mobility gap that enforces a quantization. In the present work, our main focus is to evaluate the spectral localizer for energies in the bulk of the clean system (the red intervals in Fig.~\ref{Fig:Spec_Loc_Data_size}), where there is accordingly no guarantee for a finite localizer gap. Still, we observe a robust localizer gap $g_L$ for extended bulk energy windows and a non-zero value of $Q$ inside these windows. The purpose of this appendix is to provide numerical data to substantiate this claim and to make a rigorous argument for why the spectral localizer data of the clean system should still correctly predict the quantized Hall conductance at the onset of disorder.

We calculate the localizer gap for different system sizes and find that it saturates at $N_x  = N_y = N = 15$. The data is presented in Fig.~\ref{Fig:Spec_Loc_Data_size}. We note that the sign of the topological index $Q$ obtained from the spectral localizer is opposite to that of the observed Hall conductance in the presence of disorder, which is due to a different sign convention for the Berry curvature. The spectral localizer framework takes the Berry curvature as $F_n = i (\nabla \times \bra{n} \nabla \ket{n}) \cdot \bm e_z$ (see Eq.~(1) of Ref.~\cite{Spec_loc_5_App}) as opposed to the usual convention $F_n = -i (\nabla \times \bra{n} \nabla \ket{n} ) \cdot \bm e_z$ (see chapter three of Ref.~\cite{Bernevig_App}) used in the physics context. The latter should produce an equal sign of Chern number and Hall conductance as we also derive in \eq{Eqn:Kubo_Bastin_clean}. 

We note that there is an extended energy window in Fig.~\ref{Fig:Spec_Loc_Data_size} where the localizer gap is zero. In principle, this is expected as the existence of a finite localizer gap is only guaranteed in the presence of a spectral gap \cite{Spec_loc_5_App} or a mobility gap \cite{spec_loc_mobility_gap_App}. For energies in the bulk of the clean system, there is neither a spectral nor a mobility gap and thus no guarantee for a conclusive statement on the topology of the system from the spectral localizer. However, our numerical data strongly indicates a stable localizer gap for certain energy windows in the bulk of the clean system. Assuming that this gap is indeed stable in the thermodynamic limit, the usual argument for the topological stability of the localizer gap and associated topological indices based on Weyl's inequality \cite{Spec_loc_5_App, Fine_structure_App} can be extended to the present situation. If we order the eigenvalues of two Hermitian operators $A$ and $B$ defined on an $n$-dimensional vector space as $\lambda_1  \geq \lambda_2 \geq ... \geq \lambda_n$, Weyl's inequality states that $\lambda_i(A) + \lambda_j(B) \leq  \lambda_{i + j - n}(A + B)$. By setting $j=1$ and $j=n$, it follows that $\lambda_1(A + B) - \lambda_1(B) \in [\lambda_n(B), \lambda_1(B)]$ and thus 
\begin{align}
|\lambda_i(A + B) - \lambda_i(A)| \leq ||B||_2,
\end{align}
where $||B||_2 = \text{max}\{|\lambda_1(B)|, |\lambda_n(B)|\}$ denotes the $L_2$ matrix norm, i.e. the maximum modulus of the eigenvalues. Setting $A = L^{\hat H_0}(x,y,E)$ and $B = L^{\hat H_0 + \hat W}(x,y,E) - L^{H_0}(x,y,E) = \sigma_z \otimes \hat W$, where we denote by $L^{\hat H}(x,y,E)$ the spectral localizer for a Hamiltonian $\hat H$, leads to 
\begin{align}
\left |\lambda_i\left[L^{\hat H_0 + \hat W}(x,y,E)\right] - \lambda_i \left[L^{\hat H_0}(x,y,E)\right] \right| \leq ||L^{\hat H_0 + \hat W}(x,y,E) - L^{H_0}(x,y,E)||_2 = ||\hat W ||_2.
\end{align}
This means that the change of any eigenvalue of $L^{\hat H_0 + \hat W}(x,y,E)$ relative to that of $L^{H_0}(x,y,E)$ is bounded by the maximum eigenvalue of $\hat W$. It follows for finite $g_{L^{H_0}}(E)$ that as long as $||\hat W ||_2 < g_{L^{H_0}}(E)$, the localizer gap in the bulk at energy $E$ cannot close in the presence of the disorder potential, and thus $Q$ remains unchanged. However, a finite value of the disorder strength should still lead to Anderson localization and the formation of a mobility gap in the thermodynamic limit. Hence, the topological localizer index $Q$ obtained at energies with finite $g_L$ in the clean system should agree with the expected quantization of the Hall conductance at the onset of disorder, which is what we observe. 

To conclude this section, we would like to emphasize that we do not make any analytical argument for a finite localizer gap in the absence of a spectral or mobility gap, but only observe it as a numerical fact. It remains an interesting direction for future research whether mathematically rigorous statements can be made in this regard. 

\section{Relation to ``Topological fine structure of an energy band''} 
In a recent work \cite{Fine_structure_App}, it has been observed that a trivial band that is coupled to two bands with Chern numbers $\pm 1$ can split in two non-trivial bands in the presence of disorder. This is also accompanied by a finite and non-trivial localizer gap inside the bulk energies of the clean system. However, a deeper physical reason for this remains elusive. We find that our present approach of relating the topological structure emerging from a trivial band in the presence of disorder to an energetic separation of Berry fluxes agrees well with the numerical observations made in Ref.~\cite{Fine_structure_App}.

Concretely, the model under investigation in Ref.~\cite{Fine_structure_App} is given by
\begin{align}
H(\bm k) = \begin{pmatrix} 
h_{11}(\bm k) & h_{12}(\bm k) & v \\
h_{12}(\bm k) ^*&  -h_{11}(\bm k) & 0 \\
v & 0 & 0 
\end{pmatrix} \label{Eqn:H_fine_structure}
\end{align}
with $h_{11}(\bm k) = 2(\cos(k_x) - \cos(k_y))$ and $h_{12}(\bm k) = \sqrt{2} e^{- i \pi /4} (e^{i k_x} + e^{i k_y} + i \left(e^{i (k_x + k_y)} + 1 \right)) $. The top left $2\times 2$ block describes two topological bands with Chern number $C= \pm 1$ and the term $v$ couples them to a third band centered at $E = 0$ that is trivial and completely flat for $v= 0$. A phase transition occurs at $v \approx 5.65$, where the $C = \pm 1$ bands simultaneously touch the middle band and revert to trivial bands. At small coupling $v$, transport calculations in the presence of disorder (also performed using KWANT) show that the middle band localizes while the top and bottom bands flow together and annihilate, which is the expected behavior. However, for $3.3 \lesssim v \lesssim 5.65$, the trivial middle band starts to split into two subbands that annihilate with the already present topological bands. Please see Ref.~\cite{Fine_structure_App} for data and details. 

\begin{figure}[htp]	 
{
    \vbox to 0pt {
            \raggedright
            \textcolor{white}{
                \subfloatlabel[1][Fig:Fine_structure_a]
                \subfloatlabel[2][Fig:Fine_structure_b]
                \subfloatlabel[3][Fig:Fine_structure_c]
                \subfloatlabel[4][Fig:Fine_structure_d]
            }
        }
}
{\includegraphics[trim={0.75cm 0cm 0.75cm 0.cm}, width=0.9\linewidth]{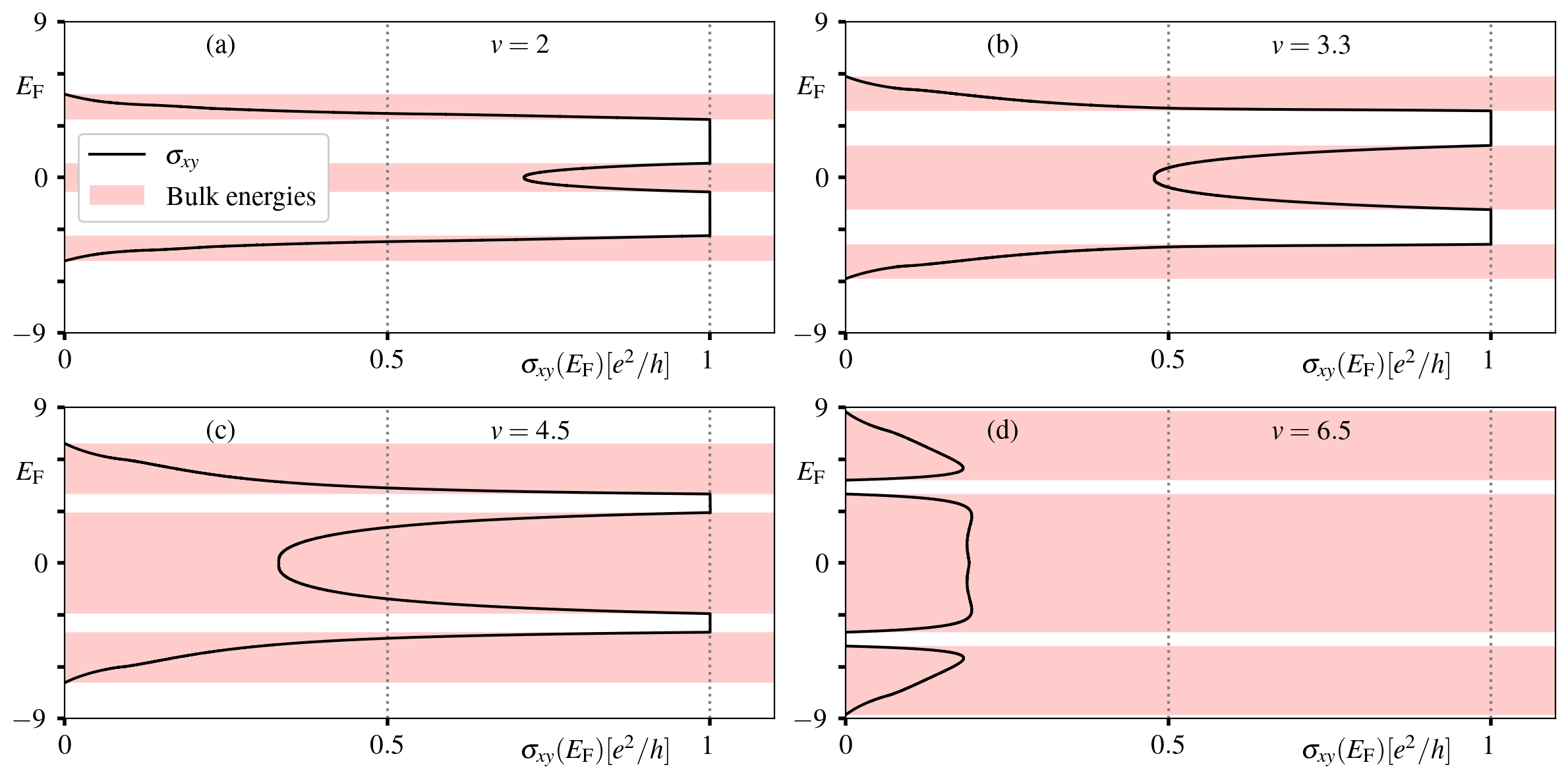}}
\caption{Clean Hall conductance $\sigma_{xy}$ of the Hamiltonian \eq{Eqn:H_fine_structure} as a function of Fermi energy $E_\mathrm{F}$ for different coupling strengths $v$. The energy windows of the bulk bands are indicated as corridors for reference. The following description of subfigures states the value of $v$ and the the behavior observed in Ref.~\cite{Fine_structure_App} for the respective value. (a) $v = 2$, $\sigma_{xy}$ does not dip below  $0.5 e^2 / h$, in the central band. The central band does not split in the presence of disorder, but simply localizes. (b) $v = 3.3$, an energy window emerges where $\sigma_{xy} < 0.5 e^2 / h$. The central band begins to split and annihilate with the upper and lower band in the presence of disorder. (c) $v = 4.5$, the window where $\sigma_{xy} < 0.5 e^2 / h$ broadens and the response to disorder is similar to $v = 3.3$. (d) $v = 6.5$, the bands have touched and a phase transition has occured such that $\sigma_{xy} < 0.5 e^2 / h$ everywhere. All bands are trivial and simply localize at the onset of disorder.
 }\label{Fig:Fine_structure}
\end{figure}

To obtain the clean Hall conductance in the sense of \eq{Eqn:Kubo_Bastin_clean} of the Hamiltonian \eq{Eqn:H_fine_structure}, an efficient algorithm for calculating the Berry curvature such as Ref.~\cite{Chern_number_numerically_App} can be used. We present the result for different values of the coupling $v$ in Fig.~\ref{Fig:Fine_structure} and find that it predicts the behavior of the system under disorder observed in Ref.~\cite{Fine_structure_App} rather well. For $v = 2$, there is no energy in the central band where $\sigma_{xy} < 0.5 e^2 / h$, as Fig.~\subref*{Fig:Fine_structure_a} shows. Consequently, one would expect that disorder simply straightens out the line and leads to one big Hall plateau, which is equivalent to completely localizing the middle band. Moving to $v = 3.3$ in Fig.~\subref*{Fig:Fine_structure_b}, an energy window opens where $\sigma_{xy} < 0.5 e^2 / h$, in which a trivial phase with $\sigma_{xy} = 0$ should form at the onset of disorder. At $v = 4.5$ in Fig.~\subref*{Fig:Fine_structure_c}, the window $\sigma_{xy} < 0.5 e^2 / h$ broadens, suggesting a similar behavior to the case of $v = 3.3$. Finally, Fig.~\subref*{Fig:Fine_structure_d} shows the case of $v = 6.5$, which is past the quantum phase transition. There, $\sigma_{xy} < 0.5 e^2 / h$ everywhere, which is consistent with a complete localization of all bands in the presence of disorder. In summary, the expectations from the evaluation of the clean Hall conductance in Fig.~\ref{Fig:Fine_structure} fully agree with the numerical results of Ref.~\cite{Fine_structure_App}, which suggests that \eq{Eqn:Kubo_Bastin_clean} is a versatile tool to gain physical insight into the topological fine structure of energy bands.

\twocolumngrid

\end{document}